\documentclass{JHEP3}

\input{epsf}
\usepackage{epsfig}
\usepackage{amssymb}
\usepackage{amsfonts}
\usepackage{amsbsy}

\def\Ab{{\bf A}}
\def\Bb{{\bf B}}
\def\Cb{{\bf C}}

\def\IC{\mathbb{C}}

\def\IZ{\mathbb{Z}}
\def\iz{\mathbb{Z}}
\def\IR{\mathbb{R}}
\def\IP{\mathbb{P}}

\def\CM {{\cal M}}
\def\CK {{\cal K}}

\def\CF {{\cal F}}
\def\CJ {{\cal J}}

\def\CO {{\cal O}}

\def\CK{{\cal K}}

\def\half{\frac{1}{2}}

\newcommand{\eq}[1]{Eq.~(\ref{eq:#1})}

\renewcommand{\Im}{{\rm Im }}
\renewcommand{\Re}{{\rm Re }}

\def\one{{\hbox{ 1\kern-.8mm l}}}

\def\tr{{\rm tr\,}}

\title{The landscape of intersecting brane models}

\author{Michael R. Douglas$^{1,\&}$ and Washington Taylor$^{2,3}$\\
$^1$NHETC and Department of Physics and Astronomy\\
Rutgers University\\
Piscataway, NJ 08855--0849, USA\\
\\
$^2$Center for Theoretical Physics\\
MIT\\
Cambridge, MA 02139, USA\\
\\
$^3$Department of Physics\\
Stanford University\\
Stanford, CA 94305--4060, USA\\
\\
$^\&$I.H.E.S., Le Bois-Marie, Bures-sur-Yvette, 91440 France\\
\\
{\tt mrd@physics.rutgers.edu, wati} {\rm at} {\tt mit.edu}
}

\abstract{We develop tools for analyzing the space of intersecting
brane models.  We apply these tools to a particular $T^6/\iz^2_2$
orientifold which has been used for model building.  We prove that
there are a finite number of intersecting brane models on this
orientifold which satisfy the Diophantine equations coming from
supersymmetry.  We give estimates for numbers of models with specific
gauge groups, which we confirm numerically.  We analyze the
distributions and correlations of intersection numbers which
characterize the numbers of generations of chiral fermions, and show
that intersection numbers are roughly independent, with a
characteristic distribution which is peaked around 0 and in which
integers with fewer divisors are mildly suppressed.

As an application, the number of models
containing a gauge group $SU(3) \times SU(2) \times U(1)$ or $SU(4)
\times SU(2) \times SU(2)$ and 3 generations of
appropriate types of chiral matter is estimated to be order ${\cal O}
(10)$, in accord with previous explicit
constructions.  As another application of the methods developed in the
paper, we construct a new pair of 3-generation $SU(4) \times SU(2)
\times SU(2)$ Pati-Salam models using intersecting branes.  

We conclude with a description of how this analysis can be generalized
to a broader class of Calabi-Yau orientifolds, and a discussion of how
the numbers of IBM's are related to numbers of stabilized vacua.}

\preprint{hep-th/0606109, MIT-CTP-3748, SU-ITP-06/15,  RU-NHETC-06-04,
NSF-KITP-06-75}

\begin{document}

\section{Introduction}

Intersecting brane models (IBM's) have been widely studied in recent
years as quasi-realistic compactifications of superstring theory
\cite{Blumenhagen:2005mu}.  One reason for their popularity is that they
appear to contain all of the necessary ingredients -- the Standard
Model, additional gauge sectors to provide supersymmetry breaking, and
the possibility of stabilizing moduli by fluxes and other means --
without requiring sophisticated mathematics to analyze.

In this work, we carry out a systematic analysis of a simple class
of models of this type, namely intersecting brane models on type IIB
orientifolds of $T^6/\IZ_2\times \IZ_2$.  
We begin with some brief comments regarding the purpose of explicit
constructions of string vacua, after which we will describe in more
detail the approach taken in this paper.

The primary goals in the study of string
compactifications at present are:
\begin{enumerate}
\item To find models which reproduce all physics observed
experimentally to date, and to work out the predictions of these
models for future observations.
\item To discover new physical mechanisms which might have observable
signatures, or might solve theoretical problems.
\item To get an overall picture of the set of all ``interesting'' models, 
and to find structure in this set
which might help in making predictions, or in uncovering deeper formal
structures in string/M theory.
\end{enumerate}
These are complementary goals, each with strengths and weaknesses.  We
note that we have no reason to think that the straightforward approach
of (1) will lead to a unique candidate vacuum, nor that any particular
mechanism found in (2) will be realized in our physical vacuum; on the
other hand (3) is at present a complicated problem which must be
simplified to make progress, running the risk of missing important
features of the physics.  In any case it seems to us that all
three goals are essential to any clear understanding of the
implications and testability of string theory.

While the methods we develop in this paper may be useful in explicit
model building as in (1), our primary interest here is more in (3).
We are interested in IBM's as a simple calculable ensemble of vacua
realizing a large variety of gauge groups and matter contents, in
which we can study the distribution of various physical features.
We follow the
point of view advocated in \cite{stat}, where the problem of
counting isolated (or ``stabilized'') vacua was set out in 
a precise way.  This work also suggested
as a simple Ansatz that various observables,
in particular intersection numbers between pairs of branes, might be
independent in ensembles of vacua coming from string theory, and this
idea was used to estimate the fraction of IBM's realizing the Standard Model
gauge group and matter representations as somewhere between $10^{-10}$
and $10^{-20}$.

In recent works of the Munich group \cite{Blumenhagen,Gmeiner} this
type of question has been analyzed in detail in the $T^6/\IZ_2\times
\IZ_2$ orientifold.  They performed an automated scan over
supersymmetric brane configurations satisfying the tadpole conditions,
and obtained statistics for the resulting ensemble of models.  
As we will review, this is
essentially a partition problem involving finding all subsets of a
fixed set of vectors which sum to a desired vector.  Such problems are
typically NP-complete, so that (for general reasons explained in
\cite{Denef:2006ad}) it is unclear that any algorithm can do this much
faster than an exhaustive search through a list of candidate models.
This requires time which grows faster than any power of the numbers
controlling the size of
the problem (Betti numbers and the numbers entering the tadpole
conditions), and indeed the computer search of \cite{Blumenhagen} ran
for about a year.

Now as long as the properties one is searching for are not too rare,
this type of intractability can be dealt with by taking a random
sample of the total ensemble, and making statistical statements based
on this sample.  But to interpret such results, it is important to
have a precise definition of the set being sampled, and an unbiased
sample.  It is not entirely clear to us whether this is true of the
algorithms used in \cite{Blumenhagen,Gmeiner}, for reasons explained
in those works and which will review below.

Nevertheless, their analysis led to some interesting observations on
the set of models.  In particular, they found that features of the
intersecting brane model field theory, such as having a U(N) subgroup
or a fixed number of generations of chiral fermions in a particular
matter sector, were largely statistically independent.  Combining the
probabilities for the gauge group and number of generations of matter
fields, in \cite{Gmeiner}, the hypothesis of independence was used to
estimate the fraction of $3$ generation Standard Models to be
$10^{-9}$ (or ``one in a billion'') in this ensemble.

In this paper, we primarily study the same ensemble of intersecting
brane models as in \cite{Blumenhagen,Gmeiner}.  Compared to these
works, there are a number of differences in our approach.  Perhaps the
main one is that we have developed algorithms which can in principle
enumerate {\bf all} configurations of a specified type, by searching
through a finite set of possibilities.  The main new ingredients which
enable us to do this are an analytic proof that the relevant brane
configurations are finite in number, and {\it a priori} bounds on the
homology classes of the branes which can appear.

This is not to say that one can usually do a complete enumeration in
practice.  For the toroidal examples we consider in detail here,
this may be possible, as the number of models is of order $10^{10}$.
However, most ensembles on more general Calabi-Yau manifolds will
have many more configurations.  Still, there are good reasons to
want algorithms which can generate the complete ensemble, such as to
be able to do unbiased random sampling.

Another difference between this work and \cite{Blumenhagen,Gmeiner} is
that we focus more closely on enumerating IBM's which contain a
specified ``visible'' gauge group and charged matter content, leaving
the rest of the model unconstrained (as was done in
\cite{Schellekens}).  We then analyze the structure of the problem
within this sector which leads to a specific dependence of the number
of models on the gauge group and matter content.  This allows
searching for Standard Models, and also for other desired structure
such as particular supersymmetry breaking gauge theories.  It also
makes the problem far more tractable computationally.  While the total
number of intersecting brane models grows exponentially in a tadpole
parameter, the number of models with fixed gauge groups grows only
polynomially.  Of course, to enumerate all models, one still needs to
consider all possibilities; we leave the question of whether the scan
over hidden sectors can be bypassed by estimating their multiplicities
for subsequent work.

A final difference is that we have not gone as far in the detailed
construction of Standard Models as \cite{Schellekens,Gmeiner}, for
example leaving explicit enumeration of ``tilted tori'' configurations
for future work.  Rather, besides methodological improvements, we
concentrate on getting detailed analytical and numerical estimates.
We find estimates for the number of models as a function of the gauge
group.  We find that the number of models with a gauge group factor
$U(N)$ goes as an inverse power of $N$.  We also investigate
intersection numbers between branes, which govern the number of matter
fields.

The overall picture which emerges from this analysis is that for this
particular orientifold, there are a calculable set of intersecting
brane models which include a wide range of specified gauge groups and
number of generations of chiral matter fields.  For the standard model
gauge group and 3 generations, the number of models is on the order of
10.  Aside from the suppression of large generations for large gauge
groups, and some mild number-theoretic enhancement of composite
intersection numbers which seems rather model dependent, there does
not seem to be any strong correlation between the specific features of
the gauge group and matter content.  Rather, this class of models seems
naturally to give rise to an ensemble of string vacua with different
gauge groups and different numbers of generations of matter fields
reasonably equally distributed.  Our results so far have also not
confirmed the interesting claim of \cite{Gmeiner} that 3 generations
(compared to 1, 2 or 4) is disfavored in this ensemble, but it may be
that we did not get far enough to see this and that it
might come out of a more complete analysis, for example including
tilted tori.

It is important to note that just because a model class does not
``explain'' or prefer the correct number of generations is not a
strong argument against it.  An arguably better application of the
statistics of vacua is to consider only vacua which fit the known data
(the Standard Model) and then look at the distribution of predictions
among these.  Interesting predictions which can be made from IBM's
include additional charged matter, particularly with exotic gauge
quantum numbers, the presence of additional gauge sectors loosely
coupled to the SM (which might be useful for supersymmetry breaking),
and hidden sectors (with purely gravitational coupling to the SM).
All of these features appear generic, and it would be very useful to
know what constraints these additional matter sectors obey.  To study
these issues it is important to have a wider class of semi-realistic
models containing at least the standard model gauge group and the
physical number of generations.  Towards this end, we include here
some discussion of how the analysis of this paper might be generalized
to a broader class of Calabi-Yau manifolds, where we might find many
more of these semi-realistic models with which to perform further
analysis.

The structure of this paper is as follows: In Section \ref{sec:group}
we study intersecting brane models on a specific toroidal orientifold
with fixed gauge group.  We prove that there are a finite number of
SUSY models with fixed tadpoles and any fixed number of brane stacks.
We develop analytic estimates of the numbers of such models, which we
check using numerical experiment.  We develop polynomial time
algorithms for generating all models with a fixed gauge group In
Section \ref{sec:intersection} we consider the intersection numbers
between brane stacks, which describe the numbers of families of matter
fields associated with strings connecting the branes.  We look at the
distribution of intersection numbers in some generic classes of
models, and analyze the statistical properties of these distributions.
In Section \ref{sec:models} we discuss the application of our general
analysis to more explicit model building, and identify a pair of new
intersecting brane models containing the standard model gauge group
and 3 generations of chiral matter.  In Section \ref{sec:notation} we
discuss the generalization of type II intersecting brane models to a
general Calabi-Yau manifold.  A similar discussion appears in
\cite{Blumenhagen:2005zh}.

One issue which we do not address in this paper is the problem of
including fluxes or other effects which stabilize the moduli.  The
intersecting brane models we consider are only isolated points in a
larger open string moduli space of supersymmetric brane
configurations.  In general, moving out into this moduli space
corresponds to recombination of branes as well as variation of their
world-volume gauge fields.  A full analysis requires understanding these
moduli spaces, and then stabilization of these moduli.
While papers such as \cite{Cascales-Uranga, Marchesano-Shiu-2} have
begun to address this question by constructing intersecting brane
models including NS-NS and R-R fluxes, a systematic and global
approach to this problem is still lacking.
In section 6, we outline a simple argument which suggests that the
number of IBM's (of a slightly more general type than we consider)
could be a good estimate for the number of stabilized vacua which
would come out of such a full analysis.  Again, we leave detailed exploration
of this idea for future work.

Our conclusions appear in section 7.

\section{Models on the torus with fixed gauge group}
\label{sec:group}

Given a particular Calabi-Yau and orientifold, there are many brane
configurations which preserve supersymmetry.  In this paper we
consider a simplification of the full classification problem for
intersecting brane models.  We assume that the gauge group is chosen
as an input, and we consider the problem of generating all SUSY
intersecting brane models with the desired gauge group.  The gauge
group can either be chosen to completely saturate the tadpole, in
which case there are no further hidden sectors of the theory, or the
gauge group can be chosen to undersaturate the tadpole, leaving room
for further ``hidden'' parts of the gauge group which would produce
the remaining contribution needed for the tadpole.  In our discussion
we use the phrase ``hidden'' sector to describe the part of the model
which does not include our gauge group of interest. We do not get into
the details of phenomenology which would be needed to make one part of
the gauge group visible in the low-energy theory.

After some initial discussion of supersymmetric branes on the toroidal
orientifold in \ref{sec:SUSY-branes}, \ref{sec:types}, in subsection
\ref{sec:symmetries} we briefly discuss useful symmetries of the
problem.  We then begin the substantive analysis in subsection
\ref{sec:finite} by demonstrating that  there are a finite number of
distinct solutions to the SUSY equations containing any fixed gauge group.
In subsection \ref{sec:estimates} we give estimates for the
numbers of brane configurations with certain properties; the results
of this somewhat detailed analysis are summarized in subsection
\ref{sec:summary}. 
In subsection \ref{sec:algorithms} we give
efficient algorithms for generating all SUSY brane models with a fixed
tadpole and fixed gauge group.  

Compared to previous work, there are two essential new ingredients in
our analysis and algorithms.  The first new ingredient comes from the
proof of finiteness, which allows us to compute explicit bounds on the
range of winding numbers of the branes which can appear.  These bounds
not only allow us to demonstrate that a given set of solutions is
complete, but also allow us to efficiently scan the space of allowed
configurations.  The second new ingredient is to enumerate brane
configurations independently of the K\"ahler moduli, and then check
for each configuration whether it is supersymmetric for some value of
the moduli; this avoids a time-consuming scan over moduli.

As we will show, the number of configurations with a fixed gauge group
grows polynomially in the total tadpole number, with a power growing
with the number of ``brane stacks'' or factors in the gauge group.
The search algorithms typically take time which is comparable to the
number of configurations; in other words one needs to search through a
number of candidates comparable to the number of solutions.  This is
essentially the same level of difficulty as for finding flux vacua, in
which the time also scales with the number of configurations.  For
example, for IIB flux vacua on a Calabi-Yau with third Betti number
$b_3$ and D3 tadpole $L$, the number of vacua grows as $(2\pi
L)^{b_3}/(b_3)!$ (for $L\gg b_3$) \cite{ad,DSZ}, and there is probably
no algorithm for finding the interesting ones which is much more
efficient than exhaustive search through a large set of candidates.
This number of flux vacua grows polynomially with a tadpole
constraint, while it grows exponentially with the number of cycles
$b_3$.  The exponential complexity of the general brane problem, on
the other hand, arises from the need to sum over all gauge groups and
thus all partitions.  While we avoid this particular exponential
complexity using the methods of this paper, it is distinct from that
which would appear on a more general Calabi-Yau with large Betti
numbers $b_i$, where again the number of vacua would grow
exponentially in $b_i$.  We discuss this issue further in Section
\ref{sec:notation}.

As we will now describe in more detail, the problem of enumerating
all intersecting brane constructions with a fixed gauge group is a
polynomial time problem, while counting and characterizing the
possible hidden sectors is in general exponential.  However, we should
emphasize that in the
$T^6/\IZ_2^2$ model, the full range of possible hidden sectors should
also be computable
due to the size of the numbers involved using the more efficient
algorithms we develop here.

\subsection{Supersymmetric branes on the toroidal orientifold}
\label{sec:SUSY-branes}

In this subsection we quickly review the problem of finding
supersymmetric brane configurations in the $T^6/\IZ_2^2$ orientifold.
We follow most closely the discussion in \cite{Blumenhagen}, but
equally good accounts include \cite{Cveticll,Blumenhagen:2005mu}.  Readers
unfamiliar with this problem can also skip ahead to section
\ref{sec:notation}, where the general setup is reviewed.

Considering $T^6$ as a product of 3 $T^2$'s with complex coordinates
$z_i$, the $\IZ_2$ actions are given by 
$(z_1, z_2, z_3)\rightarrow(-z_1, -z_2, z_3)$ and
$(z_1, z_2, z_3)\rightarrow(z_1, -z_2, -z_3)$, while the orientifold
action takes $\Omega:z_i \rightarrow \bar{z}_i$.
For the subset of branes we consider, mirror symmetry is very simple
(it amounts to a triple $T$-duality), and the IIA and IIB descriptions
are equivalent.  We are going to use a mixed IIA--IIB language,
because while the pictures are simpler in the IIA language with
wrapped D6-branes, much of the underlying mathematics and the
generalization to other Calabi-Yau orientifolds can be described most
easily in the IIB language.

In IIA language, the supersymmetric branes we consider are D$6$-branes
wrapping flat $T^3$ submanifolds of $T^6$.  Such a submanifold can be
described by a vector of six winding numbers $n_i$ and $m_i$ with
$i\in\{1,2,3\}$, defining an embedding from a $T^3$ with coordinates
$\sigma_1,\sigma_2,\sigma_3)$ to $T^6$ with coordinates $(X^i,Y^i)$ as
$$
X^i = n_i \sigma_i; \qquad Y^i = m_i \sigma_i.
$$
To get a one-to-one embedding, the 
pair of winding numbers $(n_i,m_i)$ must be
relatively prime for each $i\in\{1,2,3\}$; such a vector is referred
to as ``primitive.''
One can check that any
such submanifold is invariant under the orbifold action $\IZ_2^2$.
The orientifold $\Omega$  will act as $(n_i,m_i)\rightarrow
(n_i,-m_i)$, so a consistent brane configuration is a set of branes
which is invariant under $\Omega$.  

The mirror symmetry to IIB operates by T-dualizing the three
coordinates $Y^i$.  The mirror of such a brane then depends on the
indices $n_i$.  If all are non-zero, one gets a D$9$-brane carrying
magnetic flux determined by the indices $m_i$.  When any $n_i$ are zero,
one gets lower dimensional wrapped branes.  For now, we use
IIA notation; the relationship
between these pictures is discussed further in Section
\ref{sec:notation}.

A supersymmetric brane model consists of a set of brane stacks satisfying
a common supersymmetry condition.  Each brane stack is parameterized
by the number of branes $N$ and a set of winding numbers $n_i, m_i, i
\in\{1, 2, 3\}$.  The real part of the supersymmetry condition is
\begin{equation}
0 = m_1 m_2 m_3 - m_1 j_2 n_2 j_3 n_3 - m_2 j_3 n_3 j_1 n_1 - m_3 j_1
n_1 j_2 n_2 .
\label{eq:basic-SUSY}
\end{equation}
Here, $j_i > 0$ are the K\"ahler moduli in the IIB picture, T-dual to
complex structure moduli in the intersecting brane IIA picture.
Each brane stack has an image under $\Omega$ where $n_i' = n_i, m_i' =
-m_i$.

Using $a$ to index the distinct brane stacks, not counting images
under $\Omega$, and denoting by  $T = 8$ the bound on the total
tadpoles arising from the orientifold,
the tadpole conditions
can be written as
\begin{equation}
\sum_{a}P_a =\sum_{a}Q_a= \sum_{a}R_a= \sum_{a}S_a=  T
\label{eq:total-tadpole}
\end{equation}
where the individual tadpole contributions from each brane stack are
given by
\begin{eqnarray}
P & = &  n_1 n_2 n_3 \nonumber\\
Q & = &  -n_1 m_2m_3 \nonumber\\
R & = &  -m_1 n_2m_3 \label{eq:tadpoles}\\
S & = &  -m_1 m_2n_3 \nonumber\,.
\end{eqnarray}
With this notation, when all tadpoles are nonvanishing
the real part of the SUSY condition becomes
\begin{equation}
\frac{1}{P}  + \frac{j}{Q}  + \frac{k}{ R}  + \frac{l}{S}  = 0  \,.
\label{eq:SUSY-1}
\end{equation}
where we denote $j = j_2j_3, k = j_1 j_3, l = j_1 j_2$.  The
positivity part of the SUSY condition becomes
\begin{equation}
P + \frac{1}{j}  Q + \frac{1}{k}  R + \frac{1}{l}  S > 0 \,.
\label{eq:SUSY-2}
\end{equation}

Finally, there is a further discrete constraint from K-theory which
states that when we sum over all branes we must have
\cite{Uranga:2000xp}
\begin{equation}
\sum_{a} m_1 m_2 m_3 \equiv
\sum_{a} m_1  n_2  n_3 \equiv
\sum_{a}  n_1 m_2  n_3 \equiv
\sum_{a} n_1 n_2 m_3 \equiv
0 \; ({\rm mod} \; 2) \,.
\label{eq:K-theory}
\end{equation}

To summarize, our problem is to find all sets of ``$N_a$-stacks'' of
supersymmetric branes, where $N_a$ is a list of positive integers, for
all values of the moduli $(j,k,l)$.  A supersymmetric brane for moduli
$(j,k,l)$ is specified by a vector winding numbers $n_i^{(a)},
m_i^{(a)}$, which satisfy the SUSY conditions (\ref{eq:SUSY-1}) or
(\ref{eq:basic-SUSY}), as well as (\ref{eq:SUSY-2}) and (\ref{eq:K-theory}).  
An $N_a$-stack of branes is a set of supersymmetric branes taken with
multiplicities $N_a$, which can be completed by adjoining zero or more
additional distinct supersymmetric branes (with the same moduli) to
give a configuration which satisfies \eq{total-tadpole}.  These additional
branes will be called the ``hidden sector'' (note that they might or
might not be physically observable in a particular model).

\subsection{Types of branes}
\label{sec:types}

Our first step in solving this problem will be to classify the set of
supersymmetric branes which are compatible with the positivity
condition \eq{SUSY-2}.  This set has three components, distinguished
by the number of nonzero tadpoles, which we will call the \Ab-branes,
\Bb-branes and \Cb-branes.  They are as follows:

\begin{description}
\item {\Ab)} Four nonvanishing tadpoles:\\
In this case all $n$'s and $m$'s are nonzero.
Let us categorize the allowed sign possibilities.  First, note
that changing the signs on any set of 4 winding numbers $n_i, m_i,
n_j, m_j$ amounts to an orientation-preserving coordinate redefinition
and maps any configuration to one which is physically equivalent.
Thus, we can without loss of generality choose signs so that either
all $n$'s are positive or we have signs $+, +, -$.  
Let us first consider the case that all $n$'s are positive.  If we
choose all $m$'s positive then (\ref{eq:SUSY-1}) and (\ref{eq:SUSY-2})
cannot both be satisfied, since only $1/P$ in (\ref{eq:SUSY-1}) is
positive, and therefore $P<- Q/j$, for example, so the LHS of
(\ref{eq:SUSY-2}) would be negative.
Changing signs on all $m_i$ takes a brane to its image under $\Omega$
and leaves the tadpoles invariant, so we can assume that all branes
contributing in the sum (\ref{eq:tadpoles}) with $n_1, n_2, n_3 > 0$ have
signs for the $m_i$ of $+--, -+-$, or $--+$.  These three
possibilities give tadpole combinations with 3 positive tadpoles and
one negative, with $Q, R, S$ respectively negative for the 3 choices
of sign for the $m$'s.  Now, assume the $n$'s have the signs $++-$.
This gives a negative value for $P$, and to satisfy (\ref{eq:SUSY-1})
we must have $Q, R, S > 0$, so without loss of generality we have
signs $++-$ for the $m$'s.

To summarize, when all four tadpoles are nonvanishing, 3 must be
positive and one negative.  

\item {\Bb)} Two nonvanishing tadpoles:\\ In this case just one of
the $n_i, m_i$ vanishes.  This leads to two nonvanishing tadpoles and
two vanishing tadpoles.  A similar but simpler argument to that of
case a) shows that both nonzero tadpoles must be positive.  Various
choices of signs and which winding numbers vanish are possible, giving
all 6 possible pairs of nonvanishing tadpoles.

\item {\Cb)} One nonvanishing tadpole:\\ In this case at least two
of the $n_i, m_i$ vanish.  But if precisely two vanish and there is
one nonvanishing tadpole then there must also be precisely one
nonvanishing term on the LHS of (\ref{eq:basic-SUSY}), which is
impossible.  So three winding numbers must vanish.  In this case,
however, primitivity implies that the
nonvanishing winding numbers are all 1, so that the single
nonvanishing tadpole is 1.  Branes of this type do not contribute at
all to (\ref{eq:basic-SUSY}) and thus do not constrain the K\"ahler moduli
$j_i$.  These are referred to as ``filler'' branes in some of the
literature.
\end{description}

\subsection{Symmetries}
\label{sec:symmetries}

In this subsection we discuss the symmetries of the equations
(\ref{eq:SUSY-1}) and (\ref{eq:SUSY-2}).  We discussed above the
symmetries of the $n$'s and $m$'s which map every brane configuration
to a physically equivalent configuration.  Changing signs on all $m$'s
corresponds to switching branes under the action of the orientifold
$\Omega$.  Changing signs on a pair of $n$'s and the same pair of
$m$'s corresponds to an orientation-preserving coordinate change on
the D-branes in the system.  We use these symmetries to choose
canonical forms for the values of $n$ and $m$ in brane configurations.

There is a further set of symmetries which acts on the set of
solutions of (\ref{eq:SUSY-1}), (\ref{eq:SUSY-2}).  This set of
symmetries takes one solution to a  distinct solution with different
tadpoles.  The obvious part of this set of symmetries arises from
permutations on the 3 copies of $T^2$, associated with the 6
permutations on the indices $i$.  These symmetries also act on the
tadpoles $Q, R, S$ through the same permutation, and permute the
moduli $j, k, l$ in the same way.

This order 6
symmetry group can be extended further to include arbitrary
permutations on all four tadpoles $P, Q, R, S$.  The symmetry
operation which exchanges $P$ with one of the other four tadpoles
corresponds to 90 degree rotations on two of the tori
\begin{eqnarray}
n_i \rightarrow m_i \rightarrow -n_i &  & \\ \label{eq:p-exchange}
n_j \rightarrow m_j \rightarrow -n_j &  &  \,.\nonumber
\end{eqnarray}
To extend this symmetry to the moduli $j, k, l$, we can write these
moduli over a common denominator, $ j \rightarrow j/h, k \rightarrow
k/h, l \rightarrow l/h$, and then the permutation simply acts on the
parameters $h, j, k, l$.  When the original $j, k, l$ are rational, we
can replace the moduli with 4 integer parameters $h, j, k, l$.

This symmetry group corresponds to choices of coordinates on the
original $T^6$.  We will find it useful to use this symmetry of the
equations to simplify the discussion and analysis in several places in
the remainder of the paper.

\subsection{Proof of finiteness}
\label{sec:finite}

In this subsection we prove that the number of models on the
$T^6/\IZ_2^2$ orientifold with a fixed number of brane stacks is
finite.  This implies among other things that the number of models
with any fixed gauge group is finite.  In this subsection, and in the
following subsections where we estimate the number of models with
certain properties, we leave the total tadpole $T$ as variable, though
in the physical application for the given orbifold we have $T = 8$.

All the difficulty in proving finiteness arises from the \Ab-type branes
which can contribute negatively to \eq{total-tadpole}.  If we have
only type \Bb\ and \Cb\ branes, it is immediately clear that there are a
finite number of possible ways to saturate \eq{total-tadpole}, since
each brane stack with a winding number of absolute value $N$ or more
contributes at least $N$ to some tadpole.  Thus, we must have all
winding numbers $\leq T$, which can be done in a finite number of
ways, and we must have $\leq 4T$ brane stacks and images from that
finite set of allowed brane stacks.

The analysis with the \Ab-type branes must involve the supersymmetry
conditions in an essential way, since one can find infinite
sets of stacks by combining branes which are supersymmetric for
different moduli \cite{Kumar:2006yg}.

We clearly cannot have an infinite family with only one \Ab-type brane,
since in an infinite family some winding number must be unbounded.
This winding number contributes to two nonzero tadpoles, so with only
one \Ab-type brane it must contribute an unbounded positive amount to
some tadpole which cannot be canceled by a negative contribution.
This is a contradiction, so any infinite family must have more than
one \Ab-type brane.  
For the same reason, we cannot have an infinite family where only one
tadpole takes unbounded values.

Now, let us consider the possibility of an infinite family where two
tadpoles take unbounded values.  Without loss of generality, let's
assume these are the tadpoles $R$ and $S$.  Assume there are $r$
\Ab-type branes with tadpoles which are negative for $R$, and $s$ \Ab-type
branes with negative $S$ tadpoles.  In the first case the tadpoles can
be written $R = -a_i \nu_i, S =b_i \nu_i, i = 1, \ldots, r$ where for each
brane $\nu = |m_1|, a = n_2|m_3|,$ and $b = n_3m_2$.  In the second case the
tadpoles can be similarly written $R = a'_i \mu_i, S = -b'_i \mu_i$.
Because the other tadpoles are bounded, we have $0 \leq a_i, b_i,
a'_i, b'_i\leq B$ for each $i$, where the upper bound is $B = T^2$
unless there are other branes with negative $P$ or $Q$ tadpoles.  If
there are an infinite family of configurations with a fixed number of
branes and a fixed bound $B$, then there must also be an infinite
number with some fixed combination of values for $a_i \ldots b'_i$, so
let us take these values as fixed.  For these fixed values, the SUSY
condition for the configuration with the smallest tadpoles tells us
that there is a ratio $\lambda =l/k$ such that
\begin{equation}
b_i >\lambda  a_i, \;\;\;\;\;
a_i' > b_i'/\lambda \,.
\end{equation}
But now let us use this  $\lambda$ for the full infinite family and
compute for each member of the family the sum
\begin{equation}
\sum (b_i-\lambda a_i) \nu_i
+ \sum \lambda (a_i' -b_i'/\lambda) \mu_i
= T (\lambda + 1) \,.
\label{eq:2-bound}
\end{equation}
On the one hand, the RHS is fixed.  But on the other hand, the LHS
is a sum over linear terms in the $\nu_i, \mu_i$ with fixed positive
coefficients.  So there cannot be solutions of this equation with
arbitrarily large $\nu_i$ or $\mu_i$ and therefore there cannot be an
infinite number of configurations where two  tadpoles
are unbounded and negative.

So we see that any infinite family of brane stack configurations
solving the SUSY equations with a fixed number of stacks must have
configurations with at
least three \Ab-type branes and at least 3 of the tadpoles $Q,R, S, T$
must be negative and unbounded on a sequence of \Ab-branes.  We now
proceed to prove that this situation also cannot occur.

Let us consider the sum over all \Ab-type branes
\begin{equation}
\sum_{a} P_a
+\frac{1}{j}  \sum_a{Q_a}
+\frac{1}{k}  \sum_a{R_a}
+\frac{1}{l}  \sum_a{S_a} 
\leq T (1 +  \frac{1}{j}  + \frac{1}{k} + \frac{1}{l}  ) \,.
\label{eq:limited-sum}
\end{equation}

For each $a$, there is one negative tadpole, so the contribution for
that a to the sum can be written using the equation (\ref{eq:SUSY-1})
as, for example,
\begin{equation}
\left[ P_a + \frac{Q_a}{j}  + \frac{R_a}{ k}  -\frac{1}{\frac{1}{P_a} +
    \frac{j}{Q_a}  + 
    \frac{ k}{R_a}}  \right] + \cdots
\end{equation}
when $S_a < 0$
and similar expressions when one of the other tadpoles is negative.

But we can then use the elementary inequality
\begin{equation}
\frac{1}{x}  + \frac{1}{y}  + \frac{1}{z} 
> \frac{3}{x + y + z} 
\end{equation}
to show that
\begin{equation}
\sum_{a+} P_a
+\frac{1}{j}  \sum_{a+}{Q_a}
+\frac{1}{k}  \sum_{a+}{R_a}
+\frac{1}{l}  \sum_{a+}{S_a} 
\leq \frac{3}{2}  T (1 +  \frac{1}{j}  + \frac{1}{k} + \frac{1}{l}  ) \,,
\label{eq:limited-sum-32}
\end{equation}
where for each tadpole we only sum over positive contributions to that
tadpole.

Now, using the symmetries discussed in the previous subsection, let us
assume without loss of generality that $1 \le j \le k \le l$.  We then have
an upper bound
\begin{equation}
\sum_{a+}P_a \leq 6T \,.
\label{eq:p-bound}
\end{equation}
So we see that the sum of all positive $P$ tadpoles is bounded above
and therefore also that the negative tadpoles are bounded below by a
limit proportional to the total tadpole number $T$.

We can now repeat this argument for $Q$.  We have
\begin{equation}
\frac{1}{j}  \sum_a{Q_a}
+\frac{1}{k}  \sum_a{R_a}
+\frac{1}{l}  \sum_a{S_a} 
\leq T ( \frac{1}{j}  + \frac{1}{k} + \frac{1}{l}  ) \,.
\label{eq:limited-sum-2}
\end{equation}
Using the inequality
\begin{equation}
\frac{1}{x}  + \frac{1}{y}
> \frac{2}{x + y}
\end{equation}
we then have
\begin{equation}
\frac{1}{j}  \sum_{a+}{Q_a}
+\frac{1}{k}  \sum_{a+}{R_a}
+\frac{1}{l}  \sum_{a+}{S_a} 
\leq 2  T (\frac{1}{j}  + \frac{1}{k} + \frac{1}{l}  ) \,,
\end{equation}
so again
\begin{equation}
\sum_{a+}Q_a \leq 6T \,.
\label{eq:q-bound}
\end{equation}

Thus, we see that two of the tadpoles are bounded in any infinite
family, so there cannot be any infinite family where 3 tadpoles become
unbounded.  But the first part of the argument then shows that there
cannot be any infinite families, since there cannot be any infinite
families with only two unbounded tadpoles.

Combining these arguments, we see that any class of configuration with
a definite number of stacks and thus a definite number of factors in
the gauge group, contains finitely many configurations.  To complete
the argument that the total number of brane configurations is finite,
we need to see that the number of factors in the gauge group is also
bounded.  This follows because the bounds we just obtained on the
winding numbers do not depend on the number of stacks (beyond three).
Thus the number of distinct supersymmetric branes is finite, and since
each stack must contain a different supersymmetric brane, the number
of possible stacks is finite.

\subsection{Estimates}
\label{sec:estimates}

We can now use the analysis of the previous subsection to develop
estimates for the numbers of brane configurations containing a
particular gauge group and algorithms for enumerating these
configurations.

First, because of the existence of the filler (type \Cb) branes, any
configuration which {\it undersaturates} each of the total tadpole
constraints \eq{total-tadpole}, can be completed to a configuration
satisfying \eq{total-tadpole} by adding \Cb-branes, in a unique way.
Thus, the bulk of the problem amounts to enumerating $N_a$-stacks of \Ab-
and \Bb-type branes satisfying
\begin{equation}
\sum_{a}P_a \le T; \qquad
\sum_{a}Q_a \le T; \qquad
\sum_{a}R_a \le T; \qquad
\sum_{a}S_a \le  T.
\label{eq:under-tadpole}
\end{equation}

In general, there are other configurations as well.
First, part or all of the
specified gauge group can be realized by \Cb-type branes.  
These configurations are easy to get (of course one should keep in mind
that adding further \Cb-branes to saturate the tadpole will change the
gauge group).
Second, some configurations which violate \eq{under-tadpole} can be
completed to satisfy \eq{total-tadpole}, by adding hidden sector
\Ab-branes.  We will neglect this possibility to start, and discuss
\Ab-type branes in the hidden sector later.

As we discussed earlier, there are straightforward bounds on the types
and numbers of \Bb- and \Cb-type branes which can add up to a specified
set of tadpoles.  The difficulty is with the \Ab-branes.  One way to
deal with them, which was taken in \cite{Gmeiner}, is to scan over
values of the moduli $(j,k,l)$.  It follows immediately from
(\ref{eq:SUSY-2}) that the set of
supersymmetric branes for any fixed moduli is finite.  The problem
with this approach is that one must then scan over all values of the
moduli.
Since the moduli need not be integers, this scan is far more
difficult than the original problem.  One might bring in
number-theoretic arguments to bound the heights of the moduli, along
the lines of \cite{DeWolfe:2004ns}, but this does not look easy.

A different approach, which we will follow here, is to find {\it a
priori} bounds on the numbers and types of \Ab-branes which can appear,
which are independent of the moduli.  We then enumerate all
$N_a$-stacks, and then check for each whether all branes can
simultaneously satisfy the condition \eq{basic-SUSY} (note that we
already found the general solutions to \eq{SUSY-2}).  Since the
equations \eq{SUSY-1} for \Ab-branes are linear in the moduli $(j,k,l)$,
as are \eq{basic-SUSY} for \Bb-branes, finding the moduli which solve
them is an easy problem in linear algebra.  We then keep only the
configurations for which $(j,k,l)$ are all positive.

This is a good approach if the fraction of all brane configurations
(satisfying our {\it a priori} bounds) which turn out to be
supersymmetric is not too small.  For models with up to three stacks
on $T^6/\IZ_2^2$, the equations \eq{basic-SUSY} will always have
solutions, of which a large fraction (more than $1/2^3$) turn out to
be acceptable, so this approach works well. 

On the other hand, if the number of brane stacks is greater than $b_2$
(the number of moduli, here $3$), then the equations \eq{basic-SUSY} are
overdetermined, and this will be a problem.  In this case, one could
try a hybrid approach, in which one singles out a subset of $b_2$
brane stacks, enumerates all of these, and then completes each of
these configurations by adjoining further brane stacks chosen to
respect supersymmetry at the same (known) values of the moduli.
However we will not need this approach here.

We now proceed by systematically analyzing the classes of models with
one, two, and three stacks which undersaturate all tadpoles,
discussing as we progress the effects of increasing numbers of \Ab-type
branes on the statistics.  The upshot of this analysis is summarized
in the last subsection of this section.  Basically, we will find that
when considering configurations of up to 3 brane stacks, a stack of
$N_b$ \Bb-branes can be included in $T^4/N_b^2$ ways, while a stack of
$N_a$ \Ab-branes can be included in $T^3/N_a^3$ ways.  The reader not
interested in the detailed arguments may wish to skip directly to
subsection \ref{sec:summary}.

\subsubsection{One-stack models}
\label{sec:one-stack}

We begin our analysis by looking at how many different individual
stacks of $N$ branes undersaturate all tadpoles.  As noted above, this
does {\it not} tell us about all possible individual brane stacks
which might appear as part of a model, due to the possibility of
hidden \Ab-type branes.  This analysis, however, will get us going and
indicate the nature of the problem we are considering.  (Note that
this does, however, give all possible individual brane
stacks which can appear where all possible hidden-sector branes are
type \Bb\ or filler type \Cb\ branes.)  We begin by considering single
branes (stacks with $N = 1$), and at the end of subsection describe
the changes needed to incorporate larger values of $N$.

As discussed in \ref{sec:symmetries} there are symmetries which
permute the four tadpoles $P, Q, R, S$.  Thus, we can simplify our
analysis by putting the models we are interested in in canonical
form.  For a single brane this is particularly 
simple.  We consider the three types of brane in turn
\vspace{0.1in}

\noindent {\bf C}:
All possible
type \Cb\ branes are equivalent to the brane
\begin{eqnarray}
(P, Q, R, S) & = &  (1, 0, 0, 0)\\
(n_1, n_2, n_3) = (1, 1, 1) &  & 
(m_1, m_2, m_3) = (0, 0, 0)\,. \nonumber
\end{eqnarray}
So up to symmetries there is a single type \Cb\ brane.
\vspace{0.1in}

\noindent {\bf B}:
All type  B branes can be put in the form
\begin{eqnarray}
(0, 0, R, S) & = &  (0, 0, pr, qs) \label{eq:simple-b}\\
(n_1, n_2, n_3) = (0, p, q) &  & 
(m_1, m_2, m_3) = (1, -r, -s)\,. \nonumber
\end{eqnarray}
where $R \leq S$ and $p,q, r,
s > 0$.
 (when $R = S$ we impose $p \leq q$.)
The number of solutions of (\ref{eq:simple-b}) in integers $p, q, r,
s$ with $R \leq S \leq T$ is approximately given by
\begin{equation}
 \sum_{S \leq T} \sum_{R \leq S}d (R) d (S) 
\sim \frac{1}{2} T^2 (\ln T)^2
\label{eq:bound-b}
\end{equation}
where $d (N)$ is the number of divisors of $N$, and we have used
$\sum_{N} d (N) \sim N \ln N$.
The symmetry exchanging $P$ and $Q$ is given by (\ref{eq:p-exchange})
with $i, j = 2, 3$; this exchanges $p$ with $r$ and $q$ with $s$ along
with some sign changes.  Choosing a canonical form for each allowed
brane reduces the number of distinct branes by another factor of 2.
Thus, (\ref{eq:bound-b}) gives a bound on
the rate of growth of the number of individual type \Bb\ branes which
undersaturate all tadpoles
\begin{equation}
{\cal N}_b (T) \lessapprox  \frac{1}{4} T^2 (\ln T)^2\,.
\label{eq:approximate-bound-b}
\end{equation}
Because of the primitivity condition, we must impose the further
condition $(p, q) = (r, s) = 1$.
This modifies the sum (\ref{eq:bound-b}) to
\begin{equation}
{\cal N}_b (T) \sim\frac{1}{4} \sum_{r, s \leq T}
\frac{T}{r}  \frac{T}{s}  \frac{\phi (r)}{r}  \frac{\phi (s)}{s} 
\sim \frac{T^2 (\ln T)^2}{4 \zeta (2)^2} 
=\frac{1}{4} \left(\frac{6 }{ \pi^2} \right)^2  T^2 (\ln T)^2
\label{eq:estimate-b}
\end{equation}
where $\phi (n)$ is the Euler totient function giving the number of
integers $< n$ which are relatively prime to $n$ and $\zeta (s)$ is
the zeta function $\zeta (s) = \sum n^{-s} $.  
A brief summary of the
relevant features of the totient and zeta functions are given in
Appendix A.  
The zeta function changes the
overall constant to include the relative primality condition when
summing over powers of various sets of integers, but does not change
the scaling of the overall sum.  
\FIGURE{
\epsfig{file=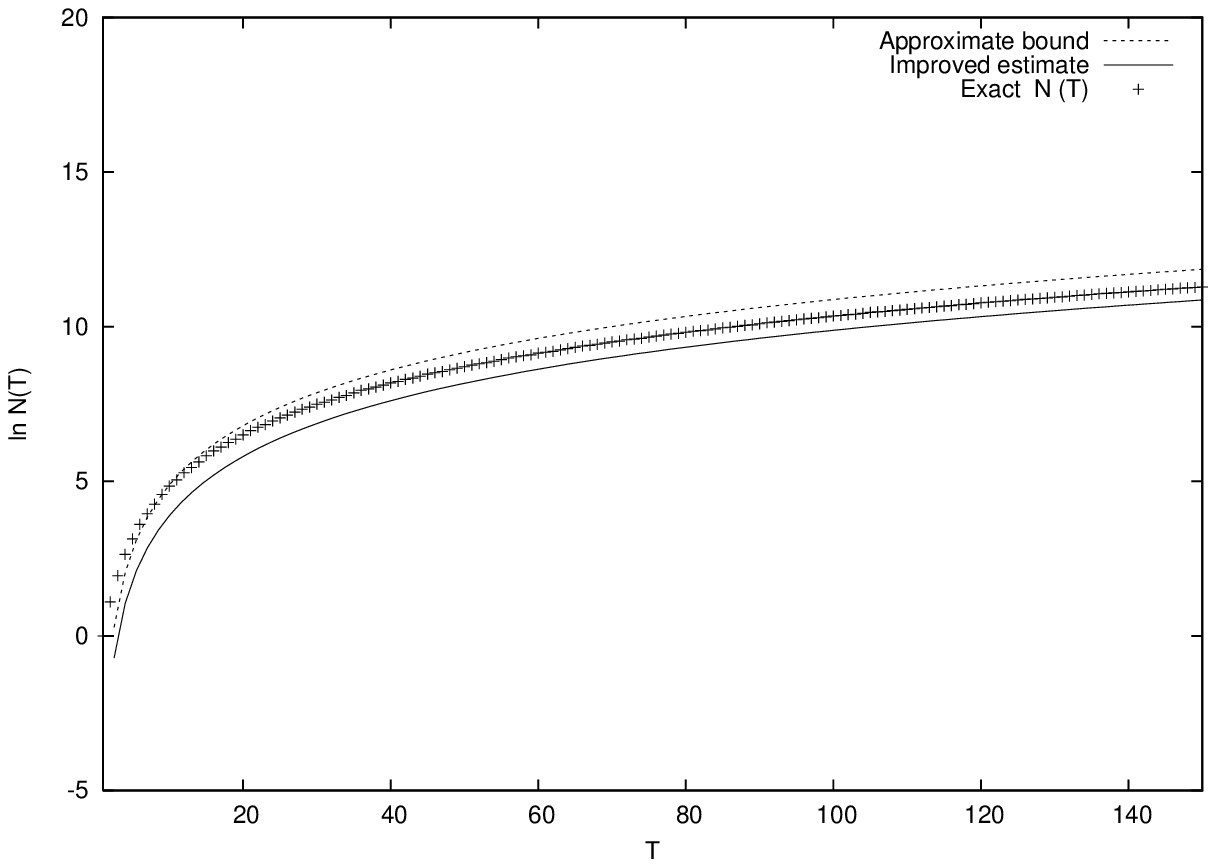,width=15cm}
\caption{\footnotesize (Log of) number of type \Bb\ branes for varying T}
\label{f:1-b}
} 
In Figure~\ref{f:1-b} we plot the (log of the) upper bound
(\ref{eq:approximate-bound-b}) and the improved estimate
(\ref{eq:estimate-b}) against the exact number of type \Bb\ branes for
general values of $T$.  At $T = 8$ the bound gives $69.19$, while the
precise number of type b branes is ${\cal N}_b (8) =71$, and the
estimate (\ref{eq:estimate-b}) gives $42.06$.  Note that for small
values of $T$ the approximate bound (\ref{eq:approximate-bound-b}) and
the improved estimate (\ref{eq:estimate-b}) are exceeded because of
configurations such as those with $R = S, p = q$, which are unchanged
under some of the permutation symmetries.  The estimate
(\ref{eq:estimate-b}) is also poor for small $T$, since $\sum_{n < N}
\phi (n)/n^2 \rightarrow \ln N/\zeta (2)$ converges to the asymptotic
form slowly.  Thus, we see that for type \Bb\ branes the precise
asymptotic form including number theoretic features is actually less
reliable at the small physical value $T = 8$ than the rough estimate
which ignores number theoretic subtleties.  We continue to fix
coefficients in this section including relative primality constraints,
but in subsequent sections we will primarily concern ourselves with the
main power law scaling of different kinds of branes.
\vspace{0.1in}

\noindent {\bf A}:
Now, let us consider type \Ab\ branes.  These can always be put in the
form
$(-P, Q, R, S)$ for positive $P$ and $0 < Q \leq R \leq S \leq T$.
(As above, when tadpoles are equal we order on the corresponding
$n$'s, and if these are equal we order on $m$ to put the branes in
canonical order; we choose all $n, m$ positive except $n_3, m_3$ as in
the discussion of Section \ref{sec:types}.)
The tadpole constraint puts an upper bound on the positive tadpoles
$Q, R, S$.  Given values of the $m$'s, we have the constraints
\begin{equation}
|n_i | \leq \frac{T}{ \prod_{j \neq i} | m_j | } 
\label{eq:n-constraints}
\end{equation}
This gives a simple upper bound on the number of type \Ab\ branes which
undersaturate all tadpoles
\begin{equation}
{\cal N}_a (T) \; \lessapprox   \;
\frac{1}{6} \sum_{| m_i | \leq T}\frac{ T^3}{ \prod {| m_i |^2} } 
\;\lessapprox \; \frac{1}{6} (\frac{\pi^2}{ 6} )^3 T^3
\label{eq:bound-a}
\end{equation}
Again, the actual number is smaller than the bound, even
asymptotically, because of the primitivity conditions $(n_i, m_i) =
1$.  The correction at large $T$ to the upper bound (\ref{eq:bound-a})
comes only from the relatively prime condition.  This correction can
be incorporated by replacing
\begin{equation}
\frac{\pi^2}{ 6}  =
\sum_{n} \frac{1}{n^2} \; \rightarrow \;
\sum_{n}\frac{\phi (n)}{n^3}  = \frac{\pi^2}{ 6 \zeta (3)} 
\label{eq:estimate-a}
\end{equation}
We graph the actual number of type \Ab\ branes which satisfy the
primitivity condition against the upper bound (\ref{eq:bound-a}) and
the estimate (\ref{eq:estimate-a}) in Figure~\ref{f:1-a}.  
\FIGURE{
\epsfig{file=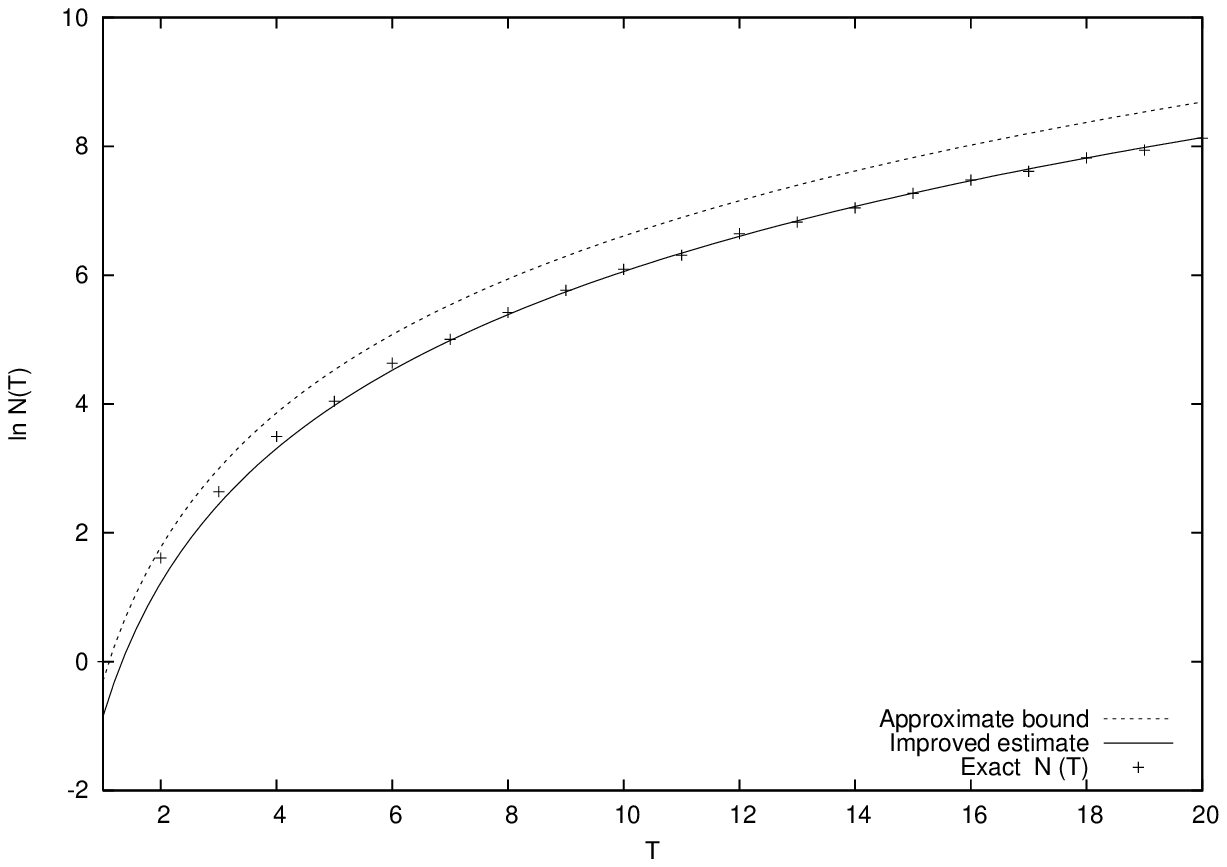,width=15cm}
\caption{\footnotesize (Log of) number of type \Ab\ branes for varying T}
\label{f:1-a}
} 
For $T =
8$, the bound gives $379.81$, and the estimate gives
$218.67$,
while the actual number of
configurations is $226$.
In this case, because the sum $\sum 1/n^2$ converges quickly, the
number theoretic corrections are correctly incorporated at a much
smaller value of $T$ so that the estimate including these corrections
is very close. 
Again, the approximate bound and the estimate are exceeded for
small $T$ because of symmetric configurations (like $n, m = (1, k, k),
(1, -1, -1)$); in this case, however, the bound rapidly exceeds the
actual number even for small $T$ because the sum in (\ref{eq:bound-a})
is finite for finite $T$, and because for small finite $T$ the number
of allowed $n_i$ is given by $\lfloor T/(m_jm_k)\rfloor$, which is
generally smaller than $T/(m_jm_k)$.

We have now described in a systematic way how to estimate the number
of single branes which undersaturate all the tadpole constraints, for
each type of brane which can occur in this orientifold.  To generalize
the discussion to a stack of $N$ branes, we simply take the formulae
we have found above and replace $T \rightarrow T/N$, since the
tadpoles from $N$ copies of a given brane $B$ are just $N$ times the
tadpoles of $B$.

Thus, we see that the number of stacks of $N$ type \Bb\ branes which
undersaturate all tadpoles goes as
\begin{equation}
{\cal N}_{Nb} \sim  \frac{9 T^2 (\ln T/N)^2}{\pi^4 N^2} 
\end{equation}
while the number of stacks of $N$ type \Ab\ branes which undersaturate
all tadpoles goes as
\begin{equation}
{\cal N}_{Na} \sim \frac{\pi^6 T^3}{6^4 (\zeta (3)^3 N^3}  \,.
\label{eq:na}
\end{equation}
Even for $N = 1, 2$, with $T = 8$
these numbers are quite small.  For example, with
$T = 8, N = 2$ we have ${\cal N}_{2a} =33$, which is close to $1/2^3$
of the 226 found at $N = 1$, and reasonably close to the estimate
(\ref{eq:na}), which gives $27.33$, despite the small numbers
involved.  As discussed above, however, the presence of more type \Ab\ 
branes introduces more negative tadpoles and makes it possible for
many other individual branes to appear as part of combinations.  The
analysis in the case of multiple stacks can carried out in a very
similar way to the analysis of this subsection, though the details are
more subtle.  We describe multiple stacks in more detail in the
following subsections.

\subsubsection{Two brane stacks}
\label{sec:two-stacks}

We now consider the case of two distinct brane stacks.  In this
section we will just be concerned with determining the power in $T$
with which different brane combinations scale, dropping constant and
log factors.  These factors could be included by a more careful
analysis.  As in \ref{sec:one-stack}, we begin by considering 
individual branes and then consider stacks of multiple branes.
\vspace{0.1in}

\noindent {\Ab\Cb, \Bb\Cb, \Cb\Cb}:
The case where one brane is type \Cb\ is straightforward, as type 
\Cb\ branes do not constrain the moduli and contribute to only a single
tadpole.  Up to symmetries, a pair of type \Cb\ branes can appear in two
ways, either contributing to the same tadpole or different tadpoles.
Thus, the number of CC type branes is a constant
\begin{equation}
{\cal N}_{cc} (T) = 2 \,.
\label{eq:cc}
\end{equation}
Given a type \Ab\  or type \Bb\ brane, generically all symmetries are broken,
and generically tadpoles are not saturated,
so there are simply 4 ways to add a type \Cb\ brane.  So asymptotically
the number of \Ab\Cb\  and \Bb\Cb\ combinations 
just go as 4 times the number of
individual \Ab\ and \Bb\ branes.  Dropping constants and log factors,
\begin{equation}
{\cal N}_{bc} (T) \sim {\cal O} (T^2) 
\label{eq:bc}
\end{equation}
and
\begin{equation}
{\cal N}_{ac} (T) \sim{\cal O} (T^3) \,.
\label{eq:ac}
\end{equation}

If we consider a stack of $N$ type \Cb\ branes and a stack of $M$ type 
\Bb\ branes, there is no constraint on the type \Bb\ branes allowed if the
stacks contribute to different tadpoles.  So the number of $M$b $+ N$
c combinations goes as (assuming $M, N \leq T$)
\begin{equation}
{\cal N}_{Mb + Nc} (T) \sim{\cal O}\left(\frac{T^2}{M^2} \right)\,.
\label{eq:mbnc}
\end{equation}
The situation is slightly more interesting for a combination of type \Ab\ 
and type \Cb\ branes.  As we discussed above, a generic stack of $M$ type
\Ab\ branes has tadpoles which scale as
\begin{equation}
M (-P, Q, R, S)\sim
(-T^3/M^2, T, T, T) \,,
\label{eq:generic-ma}
\end{equation}
so we can combine this with a stack of $N$ type \Cb\ branes for any $ N
\leq T^3/ M^2$.  Thus, there are order
\begin{equation}
{\cal N}_{Ma + Nc} (T) \sim{\cal O}\left(\frac{T^3}{M^3}\right)
\label{eq:manc}
\end{equation}
combinations of $M$ \Ab\ branes and $N$ \Cb\ branes, with $N$ allowed up to
$T^3/M^2$.
\vspace{0.1in}

\noindent {\Bb\Bb}:
A pair of \Bb-type branes is straightforward.  
Two \Bb-type branes can appear with the following four combinations of
nonvanishing tadpoles, up to symmetries
\begin{eqnarray}
(P, Q, 0, 0) & + & (P', Q', 0, 0) \\
(P, Q, 0, 0) & + & (P', 0, R', 0) \\
(P, Q, 0, 0) & + & (P', 0, 0, S') \\
(P, Q, 0, 0) & + & (0, 0, R', S')  \,.   \label{eq:bb3}
\end{eqnarray}
On the one hand,
any combination of \Bb-type
branes which can appear individually can be combined in the form (\ref{eq:bb3})
so the number of \Bb\Bb\ combinations is at least as large as the square of
(\ref{eq:estimate-b}).  But on the other hand, with any of these
combinations, each
individual brane must be one we considered in the group estimated in
(\ref{eq:estimate-b}), so the total number of \Bb\Bb\ combinations is
bounded above by a small constant factor times (\ref{eq:estimate-b})
squared.  Thus,
\begin{equation}
{\cal N}_{bb} (T) \sim{\cal O} (T^4) \,.
\label{eq:bb}
\end{equation}
It is actually straightforward to show that, more precisely,
${\cal N}_{bb} (T)\sim 3/(4 \zeta (2)^4)T^4 (\ln T)^4$.

Following the same analysis, the number of ways in which a pair of
stacks of $N, M$ \Bb\ type branes can be combined goes as
\begin{equation}
{\cal N}_{Mb + Nb} (T) \sim{\cal O} 
\left( \frac{T^2}{ M^2} \cdot \frac{T^2}{ N^2} \right)
   \,,
\label{eq:nbmb}
\end{equation}
where the two factors represent the ways in which the $M$ and $N$
stacks can be individually realized.

\noindent {\Ab\Bb}:
Now we consider combining an \Ab\ brane with a \Bb\ brane.  Here the story
becomes more interesting.  As in our discussion of \Ab\Cb\  combinations,
because the \Ab\ brane can have tadpoles which
scale as
\begin{equation}
(-P, Q, S, R) \sim (-T^3, T, T, T)
\end{equation}
we can include type \Bb\ branes with tadpoles of order
\begin{equation}
(P', Q', 0, 0) \sim (T^3, T) \,
\label{eq:larger-b}
\end{equation}
(and similar \Bb\ branes with nonvanishing tadpoles $P', R'$ and $P',
S'$) The analysis of the number of \Bb\ branes with tadpoles of this
order proceeds just as in the discussion leading to
(\ref{eq:approximate-bound-b}), but now the number of possible branes
goes as $T^4 (\ln T)^2$.  It is straightforward to check that for any
such \Ab\Bb\ combination there is a set of moduli $j, k, l$ which satisfy
the SUSY equations (\ref{eq:basic-SUSY}) for both branes.  The \Bb\ brane
in (\ref{eq:larger-b}) has $m_1 = 0$, so the tadpole condition
(\ref{eq:basic-SUSY}) just constrains the ratio of moduli
$k/l$. Putting the tadpoles from the \Ab\ brane into (\ref{eq:SUSY-1}) we
see that $k, l$ can always be taken to be sufficiently small that
there is a solution for $j$, for any ratio $k/l$.  A similar argument
holds for the other sets of order $T^4$ \Bb\ branes.  Thus, the number of
\Ab\Bb\ combinations goes as (dropping the logs)
\begin{equation}
{\cal N}_{ab} (T) \sim{\cal O} (T^7) \,.
\label{eq:ab}
\end{equation}
Note that there are also order $T^5$ configurations where the \Bb\ brane
contributes to two of the tadpoles $Q, S, R$, but these are subleading
and we will drop them.  

The number of ways in which stacks of $M$ type \Ab\  branes and $N$ type 
\Bb\ branes can be combined can similarly be computed to be
\begin{equation}
{\cal N}_{Ma + Nb} (T) \sim{\cal O}(\frac{T^7}{N^2 M^5})   \,,
\label{eq:manb}
\end{equation}
where the extra factor of $1/M^2$ appears as an extra suppression
factor on the number of \Bb\ branes because of the $P$ tadpole in
(\ref{eq:generic-ma}).

At this point we see explicitly the phenomenon mentioned above, that
inclusion of additional type \Ab\  branes can increase the number of ways
a given combination of stacks can be realized.  While the number of
individual type \Bb\ branes which undersaturate all tadpoles goes as
$T^2$, in the presence of a stack of type \Ab\  branes, there are $T^4$
type \Bb\ branes which may appear.
A more complete discussion of this effect is given in subsection
\ref{sec:extra-branes}.
\vspace{0.1in}

\noindent {\bf  AA}:
We now consider the case of two (stacks of) \Ab-type branes.  This case
is more complicated because each brane can have a different negative
tadpole.  (The number of combinations where both branes have the same
negative tadpole just goes as ${\cal N}_a^2 \approx T^6$.) We can, up
to symmetries, choose the two branes to have tadpoles\footnote{
In this argument only, we are using different sign conventions 
$m_i,n_i>0$ from elsewhere in the paper.}
\begin{eqnarray}
(-P, Q, R, S) & = &  ( -a \nu,b  \nu, m_1 n_2m_3, m_1 m_2 n_3)\\
(P', -Q', R', S') & = &  ( a' \nu',-b'  \nu, m_1' n_2' m_3', m_1' m_2' n_3')
\end{eqnarray}
where $a = n_2 n_3, b = m_2 m_3, \nu = n_1$ and similarly
for the primed variables.  This now fits directly into the context of
the proof of finiteness for configurations with two negative tadpoles
in (\ref{sec:finite}).  From the SUSY condition we must have
\begin{equation}
\frac{b}{a}  > \frac{b'}{ a'} 
\label{eq:baba}
\end{equation}
Because of the bound (\ref{eq:2-bound}), we must have
\begin{equation}
 (b-\lambda a) \nu
+ \lambda (a' -b'/\lambda) \nu'
\le T (\lambda + 1) 
\label{eq:2-bound-again}
\end{equation}
for any $\lambda$ such that $b/a > \lambda > b'/a'$.
Since $a, b, a', b' \leq T,$ this bounds the number of possible values
of $\nu, \nu'$ which can satisfy (\ref{eq:2-bound-again}).

Let us now compute the number of possible values of $\nu'$ and $\nu$
compatible with the SUSY conditions for a fixed set of values $a, b,
a', b'$.  The two individual tadpole conditions are
\begin{equation}
b \nu -b' \nu' \leq T, \;\;\;\;\;
a' \nu' -a \nu \leq T \,.
\label{eq:individual-tadpole}
\end{equation}
Because $a'/a > b'/b$,
along with $\nu', \nu > 0$  these inequalities
define a convex quadrilateral region in the $\nu', \nu$ plane whose
boundary has vertices
\begin{eqnarray}
(0, 0) & \hspace{0.8in} &  (T/a', 0)
\label{eq:4-figure}\\
(0, T/b) & & \left( \frac{a + b}{a' b- ab'} T,
 \frac{a' + b'}{a' b- ab'} T
\right) \,. \nonumber
\end{eqnarray}
We can use this quadrilateral both to estimate the number of allowed
\Ab\Ab\ brane combinations, and to determine the largest tadpoles allowed
for $P, Q'$.

Now, let us consider the ways in which $a, b, a', b'$ can be chosen.
Choosing $m_1 > 1$ essentially reduces $T \rightarrow T/m_1$ in the
selection of $a, b, \ldots$ which leads to suppression by a factor of
$m_1^2$, so the most general configurations will have $m_1 = m_1' =1$,
with other values of $m_1, m_1'$ just giving rise to an overall
constant factor of $\zeta ( 2)^2$.  We will not be extremely careful
about constant terms but will just keep around the obvious ones to get
a heuristic picture of the overall constant coefficient.  Next, we can
sum over all $n_2, n_3 \leq T$ and then all $m_{2} \leq T/n_3$ and
$m_3 \leq T/n_2$,
and similarly for the primed variables.  This will lead to on the
order of $T^4 (\ln T)^4$ possibilities.  The constraint
(\ref{eq:baba}) gives another constant factor of 1/2.

At this point let us consider three cases:\\

\noindent
{\sl i}) $b/a > b'/a' > 1$\\
{\sl ii}) $b/a > 1 >b'/a'$\\
{\sl iii}) $1 >b/a > b'/a'$
\vspace{0.1in}

In each of these cases, $a' b > ab'$.  Let us first assume that the
difference is of the same order as the larger term, $a' b-ab' \sim a'
b$, so that we can drop the smaller term in computations.  In this
approximation,
we find the following leading terms for the tadpoles of the two
branes in the three cases:
\vspace{0.1in}

\noindent {\sl i})
Here 
\begin{equation}
P = a \nu \sim \frac{ab'}{ a' b}T< T, \;\;\;\;\;
Q = b \nu \sim\frac{b'}{ a'} \leq T^2
\end{equation}
where we have used $b'a'\leq T^2$.  Similarly for $P', Q'$, so the
largest tadpoles possible in this case are
\begin{eqnarray}
(-P, Q, R, S) & \sim &  (-T ,
  T^3, T, T )\label{eq:case1}\\
(P', -Q', R', S') & \sim &  (T ,  -T^3, T, T) \,. \nonumber
\end{eqnarray}

\noindent {\sl ii})
A similar analysis shows that the largest tadpoles possible here are
\begin{eqnarray}
(-P, Q, R, S) & \sim &  (-T ,
  T, T, T )\\
(P', -Q', R', S') & \sim &  (T ,  -T, T, T) \,. \nonumber
\end{eqnarray}

\noindent {\sl iii})
In this case the largest tadpoles possible  are
\begin{eqnarray}
(-P, Q, R, S) & \sim &  (-T^3 ,
  T, T, T ) \label{eq:case3}\\
(P', -Q', R', S') & \sim &  (T^3 ,  -T, T, T) \,. \nonumber
\end{eqnarray}

In this analysis we have assumed that $a' b-ab' \sim a' b$, dropping
the second term.
The exact area of the quadrilateral (\ref{eq:4-figure}) is
\begin{equation}
\frac{1}{a' b-ab'}  ( 1 + \frac{ a}{2 b} + \frac{b'}{ 2a'}  ) T^2.
\label{eq:area}
\end{equation}
This is a measure of the number of $\nu, \nu'$'s which are allowed for
fixed $a, b, a', b'$.  We see that in case {\sl ii} the term in
parentheses is dominated by 1, and since generically $aa' bb' \sim
T^2$ and $a' b > ab'$ the area generically in this case is order 1 or
less.  In cases {\sl i} and {\sl iii} we have a term in parentheses
which can be as large as $T^2$.  Thus, we expect of order $T^2$
possible $\nu, \nu'$'s in these cases.  These numbers of $\nu, \nu'$'s
correspond to the numbers of ways we can choose the  extra factors for
the $P, Q$ tadpoles in (\ref{eq:case1}-\ref{eq:case3}).  From this
analysis we expect cases {\sl i} and {\sl iii} to dominate, with
another constant factor of 1/2 to select these cases.

Finally, however, we must look at the denominator term $a' b-ab'$ in
(\ref{eq:area}).  Generically we expect that
this number will be distributed fairly uniformly amongst numbers of
order $T^2$.  For roughly $1/T^2$ of configurations this denominator
will take the value $1$.  For these special configurations, the range
of possible values for $\nu, \nu'$ each get an extra factor of $T^2$.
For these  configurations, we can read off the maximum values for the
tadpoles, which become in case {\sl iii}
\begin{eqnarray}
(-P, Q, R, S) & \sim &  (-T^5 ,
  T^3, T, T ) \label{eq:case3-special}\\
(P', -Q', R', S') & \sim &  (T^5 ,  -T^3, T, T) \,. \nonumber
\end{eqnarray}
Note, however, that the maximum values of these tadpoles, and the
largest value of (\ref{eq:area}) will only arise when $a, a' \sim T^2,
b, b' \sim 1$.  In this case $m_2, m_3, m_2', m_3' \sim 1$ and the log
factors are dropped when computing the number of possible $a, a'$'s,
but we probably get another log from the $b'$ in the denominator of
the $a'/2b'$ in parentheses in (\ref{eq:area}).
We thus expect that the number of configurations of this type will go as
the number of $a, a', b, b'$'s ($\sim T^4$), multiplied by the
frequency with which $a' b-ab' = 1$ ($\sim 1/T^2$), multiplied by the
number of possible $\nu, \nu'$'s ($T^4$) for an overall result of $T^6
(\ln T)^4$.  When $a' b-ab' = 2$, we have a similar phenomenon but now
the area (\ref{eq:area}) gets an additional factor of $1/2$.  Summing
over all values, we expect an additional log.  A similar number of
configurations should be possible in case {\sl i}, while the number of
configurations of case {\sl ii} is suppressed by $T^2$.

Summing up the discussion, we expect that the number of \Ab\Ab\ brane
combinations should go (up to some slop in the overall constant, and
perhaps also in the power of the log) as
\begin{equation}
{\cal N}_{aa} (T) \sim{\cal O}(T^6) \,.
\label{eq:aa}
\end{equation}
A rough estimate of the constant coefficient and powers of logs
suggests that this should go more precisely as something like
$\zeta (2)^2 (\ln T)^2/32$.  These factors should not be taken too
seriously except to note that the overall constant is small; combined
with the logs this gives a factor of roughly $0.16$ at $T = 4$.

Incorporating the dependence on stack sizes is slightly more subtle in
this case.  In the total number of combinations of $N$ \Ab-stacks of one
kind and $M$ \Ab-stacks of another kind, there may be several terms in
which have different scalings.  We should have an overall suppression
by at least $N^2 M^2$ since $ab$ is suppressed by $N^2$ and $a' b'$ is
suppressed by $M^2$.  Furthermore,
the overall power of $N, M$ in the
denominator must match the power of $T$ in the numerator, so
we have schematically
\begin{equation}
{\cal N}_{Ma + Na} (T) \sim{\cal O} 
\left(\frac{T^6}{ N^2 M^2}  P_2 (1/N, 1/M) \right)
\end{equation}
where $P_2 (1/N, 1/M)$
is a homogeneous polynomial of degree 2 in its arguments.

In the case we have considered here
the values of one negative tadpole can be as low as $-{\cal O} (T^5)$,
when the scaling of the tadpoles goes as in (\ref{eq:case3-special}).
We have checked this result numerically and found agreement; note
however that the coefficient in front of the $T^5$ is something like
1/32, since the largest possible value of $a \sim a'$ goes as $T^2/4$.  
For $T = 8$, the most negative tadpole allowed is $P =-792$.

It is also of interest to consider how large the total tadpole from
the two \Ab\Ab\ branes can be.  We note first that only one total tadpole
(at most) can be negative.  Indeed, if both were negative we would
need to have $a > a', b' > b$ which contradicts $b/a > b'/a'$.
Now let us ask how negative that negative tadpole can be.  Assume we
are in case {\sl iii}, so $a > b, a' > b'$.  Then we want a lower
bound on the (possibly negative) total tadpole $a' \nu' -a \nu$.  From
(\ref{eq:individual-tadpole}) we see that
\begin{equation}
\nu \leq \frac{T + b' \nu'}{ b}  \; \Rightarrow
\; a \nu \leq T \frac{a}{b}  + a' \nu'
\end{equation}
from which it follows
\begin{equation}
a' \nu' -a \nu \geq -T \frac{a}{b}  \geq -T^3 \,.
\end{equation}
Thus, while the individual tadpoles can scale as
(\ref{eq:case3-special}), the sum of the two tadpoles must scale as
\begin{equation}
(P' -P, Q-Q', R + R', S+S') \sim
(-T^3,T, T, T)
\label{eq:total-tadpole-scaling}
\end{equation}
or just the same as a single \Ab-type brane.  
We will find this result useful in our further analysis of more
complicated configurations.
\vspace{0.1in}

To summarize the results of our analysis, in this subsection we have
shown that the numbers of \Bb\Bb, \Ab\Bb, and \Ab\Ab\ combinations of
brane stacks which admit some common K\"ahler moduli and which
undersaturate all the tadpole conditions go as $T^4, T^7, T^6$
respectively.  We have explicitly computed the number of such
combinations up to $T = 8$ (for stacks containing a single brane).
The results of this explicit computation are shown in
Figure~\ref{f:2-combinations}.
\FIGURE{
\epsfig{file=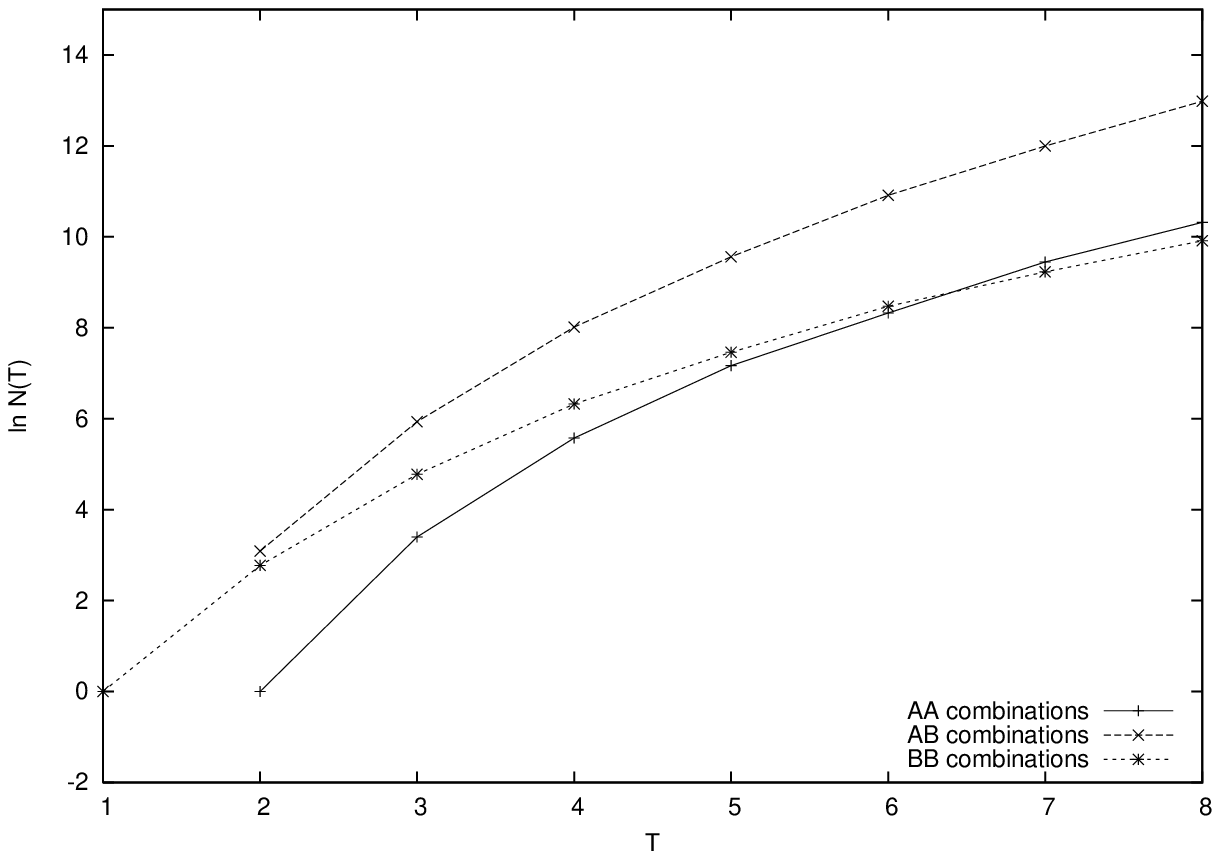,width=15cm}
\caption{\footnotesize (Log of) number of type \Ab\Ab, \Ab\Bb, 
\Bb\Bb\ branes for varying T}
\label{f:2-combinations}
} 
Note that we have connected the data points with lines for clarity,
but we have not graphed the explicit analytic predictions, which like
the case of \Bb\ branes (\ref{eq:estimate-b}) converge slowly due to the
log factors.  As predicted, the number of \Ab\Bb\ branes is the largest,
and the number of \Ab\Ab\ branes grows faster than the number of
\Bb\Bb\ branes,
but the \Ab\Ab\ and \Bb\Bb\ curves cross since the 
\Ab\Ab\ coefficient is much smaller than the \Bb\Bb\ coefficient.

At $T = 8$ we find that the exact numbers of brane combinations are
\begin{eqnarray}
{\cal N}_{aa} (8) & = &  30,255\\
{\cal N}_{ab} (8)& = & 434,775\\
{\cal N}_{bb} (8)& = & 20,244
\end{eqnarray}

The numbers at $T = 4$ are
\begin{eqnarray}
{\cal N}_{aa} (4) & = & 264\\
{\cal N}_{ab} (4)& = & 3,029\\
{\cal N}_{bb} (4)& = & 558
\end{eqnarray}
These configurations are equivalent to brane stacks with all $N = 2$
at $T = 8$, and are suppressed relative to the $N = 1, T = 8$ results
above by approximately the expected factors of $2^6, 2^7, 2^4$, though
in each case growth is slightly faster due to logs.

\subsubsection{Three stacks}
\label{sec:more-stacks}

{}From the complications in the preceding argument for the number of
two-stack \Ab\Ab\ configurations, one might worry that the problem will
become increasingly more complicated, and tadpoles will become
increasingly more negative as more type \Ab\  branes are included.
Fortunately, however, from the second part of the finiteness argument,
we know that two tadpoles can have maximum total positive contribution
of $6T$.  Thus, branes with the other two tadpoles negative are
constrained by the same type of argument as in the above \Ab\Ab\ analysis,
with the only difference being that the $a, b, \ldots$ are constrained
by a factor of $6T$ instead of $T$.  Thus, no matter how many branes
we have, the largest tadpoles can have absolute value no larger than
$T^5$ and the second largest no larger than $T^3$ as in
(\ref{eq:case3-special}).  Furthermore, when pairs of \Ab-type branes
are considered their total tadpoles must scale as
(\ref{eq:total-tadpole-scaling}), or just as a single \Ab-type brane.
This simplifies the analysis of
configurations with more branes.

In this subsection we briefly describe three-stack configurations of
various types.  We content ourselves with some simple scaling
arguments and polynomial bounds on the rate of growth of various
combinations.

Let us begin with \Bb\Bb\Bb\  combinations.  It is straightforward to see that
by choosing 3 different pairs of tadpoles for the three \Bb-type branes
we have order $T^6$ configurations.  For $N, M, L$ \Bb-type stacks, the
number of possible configurations goes as
\begin{equation}
{\cal N}_{Nb + Mb + Lb} (T) \sim{\cal O} 
\left( \frac{T^6}{ N^2 M^2 L^2} 
\right)\,.
\label{eq:bbb}
\end{equation}

Now consider \Bb\Bb\Ab.  If we take an \Ab-type brane with tadpoles of order 
 $(P, Q, R, S) \sim(-T^3, T, T, T)$, then we can consider \Bb-type branes
 with $P, Q$ and $P, R$ tadpoles.  There are $T^4$ of each type of 
 \Bb\ brane possible.  These branes fix the ratios of moduli $k/l, j/l$ and
 are compatible with any \Ab\ type brane with negative $P$ tadpole by
 just taking all of $j, k, l$ small enough.  Thus, there are $T^{11}$
 possible combinations of this type.
For 
$N, M$-stacks of \Bb\ branes and an $L$ stack of \Ab-branes, we have
\begin{equation}
{\cal N}_{N b + M  b + L a} (T) \sim{\cal O} 
\left( \frac{T^{11}}{ N^2 M^2 L^7} 
\right)\,
\label{eq:bba}
\end{equation}
where as in the \Bb\Ab\ case, the numbers of each kind of \Bb\ brane are
suppressed by an extra factor of $L^2$.

Now consider \Bb\Ab\Ab.  Since any \Ab\Ab\ combination has tadpoles which scale
at worst as (\ref{eq:total-tadpole-scaling}), or just as a single
\Ab-type brane, we again can have at most order ${\cal O} (T^4)$ choices
for the type \Bb\ brane.  Let us assume that the b brane has nonzero
tadpoles $P, Q$.  Such a brane puts a constraint on the ratio of
moduli $k/l$, but this constraint can be made compatible with any two
mutually compatible a branes with negative tadpoles in $P, Q$ by just
making $k, l$ small enough.  Thus, the number of \Bb\Ab\Ab\ branes should go
as $T^{10}$.  For a $N$-stack of \Bb\ branes and $M, L$ \Ab\ stacks, we have
\begin{equation}
{\cal N}_{N b + M a + L a} (T) \sim{\cal O} 
\left( \frac{T^{10}}{ N^2 M^2 L^2}  P_2 (1/M, 1/L)
\right)\,.
\label{eq:baa}
\end{equation}

Finally,  consider \Ab\Ab\Ab.
Since as shown in (\ref{eq:total-tadpole-scaling}), any two of the
\Ab-type branes must have a total tadpole of the same form as a single
\Ab-type brane, combining with a third \Ab-type brane seems to lead to at
most an additional factor of $T^3$.  Since not all sets of 3 \Ab-type
branes are mutually compatible with any common set of moduli, this
leads to an expected upper bound on the number of \Ab\Ab\Ab\ combinations
\begin{equation}
{\cal N}_{N a + M a + L a} (T) \sim{\cal O} 
\left( \frac{T^{9}}{N^2 M^2 L^2}  P_3 (1/N, 1/M, 1/L)
\right)\,.
\end{equation}
Because this family of configurations is subleading, and we will not
use it in any specific model-building constructions, we do not carry
the analysis of 3 \Ab-stack configurations further  here.  It should be
possible, however, to analyze this set of configurations in more
detail using the constraints from the proof of finiteness and methods
like those used in the two \Ab-stack analysis of the previous
subsection.  We leave a further analysis of this type for future work.

\subsubsection{Hidden sector \Ab-type branes}
\label{sec:extra-branes}

We have found that including \Ab-type branes can enhance the number of
ways of including brane stacks, particularly of \Bb-type branes.  One
might worry that with more and more \Ab-type branes in the hidden
sector, more and more brane combinations could be realized in the
visible sector for a given gauge group.  This does not happen,
however, essentially because of (\ref{eq:total-tadpole-scaling}).  To
see this concretely, let us return to the number of ways we can
include a $U(N)$ factor in our gauge group from $N$ type \Bb\ branes.  In
the absence of \Ab-type branes, the number of possible stacks
(\ref{eq:estimate-b}) went as $T^2/N^2$ (dropping logs).  As we found in
(\ref{eq:larger-b}), the inclusion of an \Ab-type brane in the hidden
sector increases this to order $T^4/N^2$.  From
(\ref{eq:total-tadpole-scaling}), however, we see that even when two A
branes are included in the hidden sector the number of possible \Bb-type
branes still goes as $T^4/N^2$.  In fact, when a second \Ab-type brane is
added in the hidden sector, no new type \Bb\ branes at all become
possible, since the limits on the tadpoles of the type \Bb\ brane are the
same as with a single \Ab\ brane in the hidden sector, and, as we showed
in the discussion above, all \Bb\ branes with a nonzero tadpole where the single
\Ab\ brane has a negative tadpole and
which undersaturate the total tadpole condition can be realized for
some values of the moduli.  Thus, we see that no matter how many
\Ab-type branes are included in the hidden sector, including a stack of
\Bb-type branes will always give rise to a factor of order $T^4$ in the
total number of possible models.

How about type \Ab\  branes where a type \Ab\  brane is included in the hidden sector?
From the analysis of \Ab\Ab\ combinations in \ref{sec:two-stacks} we note
that in the order $T^6$ cases where $a' b-ab' = 1$ and $ b = b' = 1,
a \sim{\cal O}
(T^2)$, the first brane can be chosen in ${\cal }(T^5)$ different
ways.  The second brane is then constrained to have $a' = a + 1$, and
$\nu'$ has a range of size $T$ from the quadrilateral
(\ref{eq:4-figure}).  So for each of the order $T^5$ choices for the first
brane there are order $T$ choices for the second brane.  This shows
that with a hidden \Ab-type brane, there can be $T^5$ \Ab-type branes in
the visible sector.  As for the \Bb\ branes, however, the presence of
another \Ab-type brane in the hidden sector cannot increase the number
of possible branes in the visible sector, since the additional hidden
brane cannot change the scaling of the tadpoles.  
By a similar argument to that above for type \Bb\ branes,
we expect that
even with an arbitrary number of \Ab-type branes in the hidden sector,
the scaling of the number of models will get a factor of at most $T^5$
from the first \Ab-stack, and at most $T^3$ from subsequent stacks,
though we do not have a complete proof of this assertion.

\subsection{Summary of results}
\label{sec:summary}

In summary, we have found that we can give an approximate asymptotic
upper bound for the number of models with any desired gauge group
appearing as a subgroup by
including a polynomial factor of
\begin{equation}
\hat{\cal N}_{N_ b  b} =\frac{T^4}{N_b^2}
\label{eq:b-factor}
\end{equation}
for each stack of $N_b$ \Bb-type branes.  This factor does not include
some additional log factors which may be particularly relevant at
small $T$ and assumes that there is at least one \Ab\ brane in the
hidden sector; if there are no \Ab\ branes in the hidden sector the
factor is reduced to $T^2/N_b^2$.  Assuming a hidden sector \Ab\
brane, a factor of
\begin{equation}
\hat{\cal N}_{N_a a}=\frac{T^5}{N_a^2}
\label{eq:a-factone}
\end{equation}
must be included for the first stack of $N_a$ \Ab-type branes; this
factor reduces to $T^3/N_a^3$ in the absence of hidden sector 
\Ab\ branes.  We are fairly sure, though we have not definitively proven,
that the contribution to the upper bound from the second and further
stacks of \Ab-type branes can be reduced to
\begin{equation}
\hat{\cal N}_{N_a a}=\frac{T^3}{N_a^2}
\label{eq:a-facttwo}
\end{equation}
whether or not there are hidden sector \Ab-type branes.
For gauge groups with many factors, the product of these
expressions will give an upper bound, but the actual number will be
fewer as the moduli will be overconstrained.  As an example, the
number of ways in which the Pati-Salam gauge group $U(4) \times U(2)
\times U(2)$ arises from \Ab\Ab\Bb\ combinations (assuming a
hidden sector \Ab\ brane) is bounded from the above estimates as
$T^{12}/2^8 \sim 3 \times 10^8$.  In practice, we expect that this is an
overestimate and that the largest number of combinations may come from
\Ab\Bb\Bb\  combinations.  We will discuss explicit model building further in
Section \ref{sec:models}.

We emphasize again that the configurations we are counting  are those
in which a desired gauge group is realized as part of a complete model
which saturates the tadpoles.  For any given gauge group which we
impose as a subgroup of the full UV gauge group arising from the
intersecting branes, each of the
polynomial number of configurations we are counting will  have a
large number of possible completions to a complete model.  Generally
the number of ``hidden sector'' completions will go roughly as
$e^{T-T_m}$ where $T_m$ characterizes the tadpole contribution from a
given model which undersaturates the complete tadpole cancellation
condition.  By analyzing the different ways in which the desired group
can be realized as a piece of an intersecting brane model, we reduce
the problem of analyzing the space of models to a polynomial problem.
We can analyze this polynomial number of solutions for configurations
of physical interest, and then in principle we could go on to do a
systematic analysis of the hidden sector possibilities in models of
particular interest.

\subsection{Algorithms}
\label{sec:algorithms}

In this section we have focused first on proving that the number of
models which contain a desired gauge group is finite, and then on
using the insight from the proof of finiteness to estimate the number
of models with a given gauge group.  We can also, however, use the
same methods used to give estimates for the number of models with
particular brane types and stack sizes to construct algorithms which
explicitly enumerate these models.  In general, these algorithms can be
extremely efficient, and have time and memory requirements roughly in
proportion to the number of solutions being constructed.

To be more explicit, consider the problem of explicitly enumerating one
of the classes of branes we have described above.  Assume first that
we have already constructed some brane configuration with total
tadpoles $P, Q, R, S \leq T$, and we wish to construct all possible
ways of including a stack of $N$ type \Bb\ branes while still
undersaturating all tadpoles.  As in the discussion proceeding
(\ref{eq:bound-b}), we are searching for sets of four integers
satisfying constraints like $pr \leq T-R, qs \leq T-S$.  By simply
looping over $p \leq T-R$ and then looping over $r
\leq\lfloor(T-R)/p\rfloor$, we can search for all allowed pairs $p, r$
in time $(T-R) \ln (T-R)$, which is roughly the number of solutions
expected, and similarly for $q, s$.  This gives us a candidate set of
allowed type \Bb\ branes to include, and we can then explicitly check the
SUSY conditions for each possibility.  This procedure can be iterated
to include an arbitrary number of \Bb-type stacks, and \Cb\ type stacks can
be included by simply checking which tadpoles are not saturated.

This leaves us with the problem of constructing algorithms for finding
combinations of stacks of type \Ab\  branes.  Using the bounds on
individual tadpoles from these branes which we found in the proof of
finiteness, we can readily construct such algorithms.  For single 
\Ab\ branes with for example a negative P tadpole, we just loop over all
possible values of the $n$'s and $m$'s which have $Q, R, S \leq T$.
As in (\ref{eq:n-constraints}), this can be done in time of order
$T^3$ and generates ${\cal O} (T^3)$ possible \Ab-type branes.  

A pair of \Ab-type branes with the same negative tadpole can be
constructed by just repeating the single \Ab\ brane procedure,
constraining the sum of the positive tadpoles.  For a pair of 
\Ab-brane stacks with different negative tadpoles, we proceed as in the 
\Ab\Ab\ analysis in subsection \ref{sec:two-stacks}.  We first loop over
possible values of $n_2, n_3, m_2, m_3$ and similar for the primed
variables.  This gives $a, b, a', b'$, which define the quadrilateral
(\ref{eq:4-figure}), which we can then scan over for all possible
values of $\nu, \nu'$.  This generates a set of candidate \Ab\Ab\ %
possibilities which we can then check for the existence of moduli
satisfying the SUSY constraints.

Constructing a set of three \Ab-type branes proceeds again in a similar
fashion.  The worst case here is when all three type \Ab\  branes have
different negative tadpoles.  Because of the bounds (\ref{eq:p-bound},
\ref{eq:q-bound}), however, we know that the sum of positive tadpoles
is bounded by $6T$ for one of the tadpoles which has a negative
contribution from one \Ab-brane.  (In fact, a more careful analysis
shows that in this 3-brane case the sum is bounded by $2T$, which
helps with constant factors but not with scaling.) From this, we can
sum for the first two \Ab-branes as in the above \Ab\Ab\ discussion, just
keeping (without loss of generality) $S_1 + S_2 \leq T, R_1 + R_2
\leq 2T$.  We construct all possible pairs under these conditions which
undersaturate the tadpoles $P, Q$ as above.  We can then enumerate all
possible third \Ab-branes given the resulting upper bounds on all
tadpoles besides $R$.

For more than three \Ab-type branes, we can generalize this analysis.
Basically the idea is that the tadpole constraints combined with
inequalities like (\ref{eq:baba}) coming from the SUSY constraints
lead to a set of inequalities like those determining the quadrilateral
(\ref{eq:4-figure}), so that we can loop over the various winding
numbers and construct a set of allowed combinations of $k$ distinct
A-brane stacks in time of order $T^{3k}$.  For $k > 3$, however, the
SUSY conditions generically will overdetermine the moduli, so this
approach becomes less and less efficient as the fraction of
SUSY-allowed combinations decreases.  To construct all configurations
with larger numbers of type \Ab\  branes, it is probably most efficient to
first construct all combinations which are large enough to uniquely
determine the moduli (which will generically be 3 \Ab\ brane stacks but
may be more in special cases).  Then, one can construct all \Ab-type
branes which are compatible with a particular set of moduli.  The
number $C_m$ of such branes must be finite for any set of moduli $m$
because of the inequality (\ref{eq:SUSY-2}) which imposes a
positive-definiteness condition on the tadpoles for fixed values of
the moduli.  If the number of stacks required to fix the moduli was
$d$, we can then check all $C_m^{k- d}$ possible ways of combining
these branes into a configuration which undersaturates all tadpoles.
This gives a complete algorithm for constructing all allowed
configurations of $k$ type \Ab\  brane stacks.

\section{Distribution of intersection numbers}
\label{sec:intersection}

Now that we have a systematic approach to estimating the number of
brane configurations realizing a desired gauge group, and efficient
algorithms for generating such configurations, we want to analyze some
further physical features of these different models.  The simplest
next feature to analyze is the intersection number between branes,
which determines the number of chiral fermions associated with strings
connecting these branes \cite{Berkooz:1996km}.  These chiral fermions
transform in the fundamental representation of the gauge group
associated with one brane and in the antifundamental representation of
the gauge group associated with the other brane.  We will not discuss
the non-chiral spectrum other than to recall that all of the branes we
are using (on the torus orientifold) have adjoint matter, and that
other vector-like matter (such as the Higgs doublets) is obtained 
by taking pairs of branes which coincide along one axis ({\it i.e.}
with the same $(n_i,m_i)$ for some $i$).

In this section we will look at the distributions of intersection
numbers for combinations of different types of branes.  There are
several motivations for considering these distributions.  Most
generally, we are interested in looking for correlations of any kind
in the landscape of string vacua.  As discussed in the introduction,
such correlations are ultimately necessary to make predictions,
assuming that all string vacua are viable.  A strong correlation, such
as an observation that all string models of this type with 3
generations of quarks also have 3 generations of leptons, would be
very suggestive and would motivate looking for similar correlations in
other regions of the landscape.  On the other hand, a smooth broad
distribution of intersection numbers with no correlations suggests
that it may be difficult to use string theory to predict the number of
generations found in nature.  As a more specific application, we can
use the distributions of intersection numbers to estimate what
fraction of the total set of models with a fixed gauge group have a
given number of generations (like 3).  We carry out such an estimate
in the next section using the results developed in this section.

To understand the correlations between different intersection numbers,
it is useful to introduce the information theory notion of {\it mutual
entropy}.  Given a probability distribution  on possible values of a
variable $x$, where $p_{i}$ is the probability that $x = i$, the
(base 2) {\it entropy} of $x$ is defined to be
\begin{equation}
H (x) = -\sum_{i}p_{i} \log_2 p_{i} \,.
\end{equation}
This quantity expresses the number of bits of information which are
needed on average to encode a given instance of the random variable
$x$.  (For example, if $x = 1, 2$ each have probability 1/4, and $x =
3$ has probability 1/2, then  $H (x) = 3/2$; if $x = 3$ we can encode
this in the bit 0, and we can encode $x = 1, 2$ in the two-bit
sequences 10, 11.)
Given a distribution on two variables $x, y$, the mutual entropy is
defined to be
\begin{equation}
{\cal I} (x, y) = H (x) + H (y) -H (x, y) \,.
\end{equation}
The mutual entropy, also called the relative entropy, encodes the
number of bits of information given about $y$ if $x$ is known (or vice
versa).  If two variables are completely independently distributed,
then ${\cal I} = 0$ (knowing $x$ gives no information about $y$).  If two
variables are completely correlated, so that there is a one-to-one
function relating values of $x$ with associated values of $y$ then
${\cal I}
(x, y) = H (x) = H (y)$.

Now, let us define the winding numbers whose distributions we are
interested in.  If we have two brane stacks, where the first has
winding numbers $(n_i, m_i)$ and the second has winding numbers
$(\hat{n}_i, \hat{m}_i)$, then the intersection number between the
branes is
\begin{equation}
I = \prod_{i}(n_i \hat{m}_i-\hat{n}_im_i) \,.
\label{eq:intersection}
\end{equation}
Because of the orientifold, there is a distinction between \Ab\ and 
\Bb\ type branes $a, b$, which have images $a' \neq a$ and $b' \neq b$
under the action of $\Omega$, and a C type brane $c$, which is taken
to itself under the action of $\Omega$.  Given two \Ab-type branes $a,
\hat{a}$, for example, the intersection numbers $I_{a \hat{a}}$ and
$I_{a\hat{a}'}$ are distinct, and must be computed separately.

\subsection{Single intersection numbers}

A coarse characterization of the distribution of individual
intersection numbers can be given by considering the standard
deviation of the set of intersection numbers within a given ensemble
of pairs of branes.  In the absence of negative tadpoles, we might
expect that typical winding numbers would be of order $T^{1/3}$,
giving typical intersection numbers of order $T^2$.  Even for type 
\Bb\ branes, however, this is not quite right.  Generally the winding
numbers are not evenly distributed.  For example, from the analysis
leading to (\ref{eq:estimate-b}), we see immediately that at least
order $T^2$ of the $T^2 (\ln T)^2$ type \Bb\ branes undersaturating all
tadpoles have three winding numbers ({\it e.g.} $m_i$) of order 1, and
two winding numbers ({\it e.g.} $n_2, n_3$) of order $T$.  Thus, there
are at least order $T^4$ \Bb\Bb\ combinations of the form $n \sim (0, T,
T), m \sim (1, 1, 1), \hat{n} \sim (1, 1, 1), \hat{m} \sim (0, T, T)$
which have $I_{b \hat{b}} \sim T^4$.  It is also easy to show that a
\Bb\Bb\ intersection number cannot be larger than $8T^4$, so we expect the
rate of growth of the standard deviation in this ensemble to be order
$T^4$ up to logs.  Because of the negative tadpoles on type \Ab\  branes,
which allow branes to have larger winding numbers, we expect a broader
distribution of intersection numbers for fixed $T$ when other brane
types are considered.  For example, in an \Ab\Bb\ configuration with a
hidden sector \Ab\ brane, we can have \Ab\ brane winding numbers which go as
$n \sim (T^3, T, -T), m \sim (1, 1, -1)$ and \Bb\ brane winding numbers
which go as $\hat{n} \sim (1, T^3, 1), \hat{m} \sim (0, 1, -T)$.  This
would lead to a reasonably generic class of configurations with
intersection numbers going as $T^7$.  We have not found any
configurations which would have faster growth, so we expect the
standard deviation of intersection numbers of pairs of branes to grow
polynomially in $T$, with a rate of growth somewhere between $T^4$ and
$T^7$.

We have computed the intersection numbers for all brane combinations
of each type (assuming no hidden sector \Ab\ branes) at various values of
$T$.  The standard deviations of the resulting distributions are
plotted in Figure~\ref{f:2-sd}.  We see that indeed the overall standard
deviation grows faster than $T^4$; the growth is dominated by \Ab\Bb\ %
configurations, which in the absence of hidden sector \Ab\ branes have
intersection numbers which grow
as $T^5$ from configurations with, for example, $n \sim (T, T, -T), m
\sim (1, 1, -1)$ and $\hat{n} \sim (1, 1, T^3), \hat{m} \sim (0, T,
-1)$.  
\FIGURE{ \epsfig{file=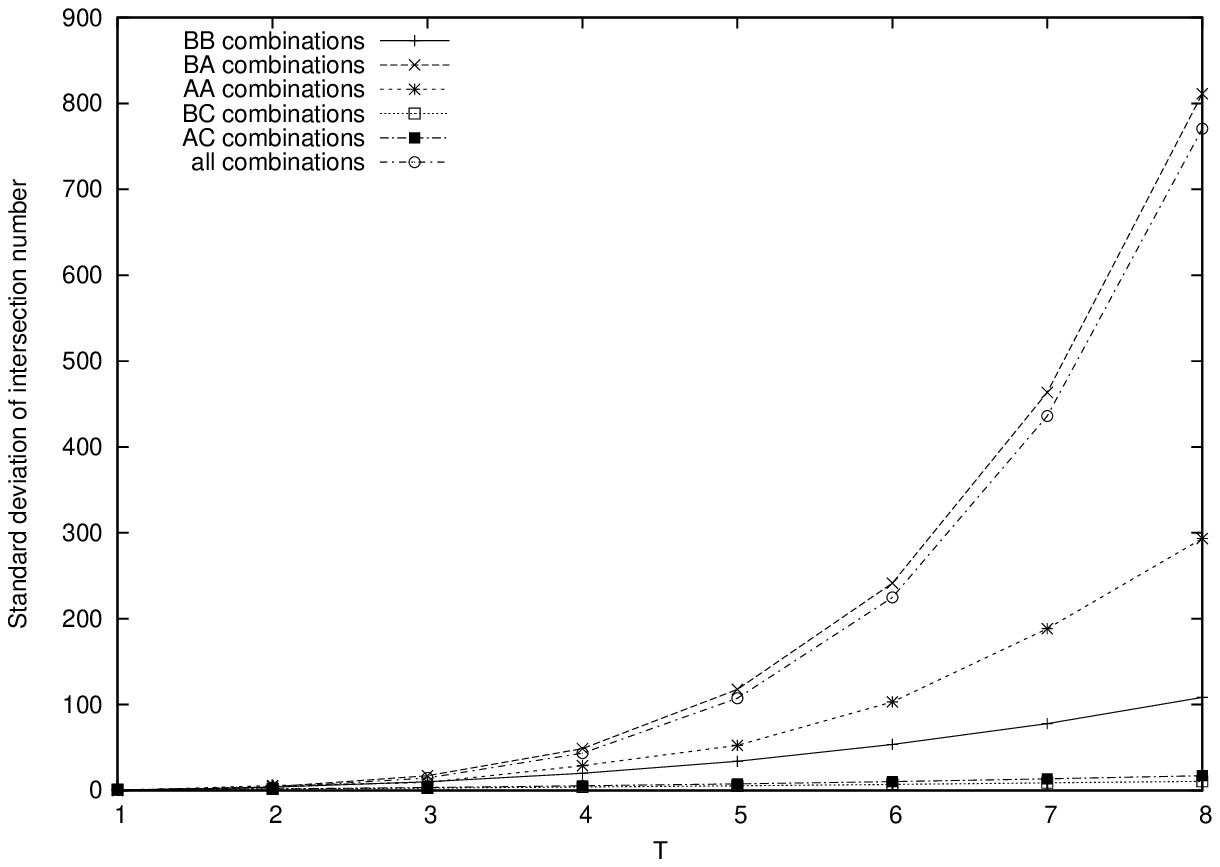,width=15cm}
\caption{\footnotesize Standard deviation in intersection number for
  different brane combinations}
\label{f:2-sd}
} 
The intersection numbers are generally much smaller for combinations
containing type \Cb\ branes, as all nonzero
winding numbers for type \Cb\ branes are 1.

The entropy $H$ of the distribution of intersection numbers characterizes
the amount of information needed to fix a single intersection number.
For a reasonably smooth distribution, generic values will appear with
frequency roughly $2^{-H}$.  We have computed the entropies of the
distributions of intersection numbers for different types of branes,
which are plotted in Figure~\ref{f:2-entropy}.
\FIGURE{
\epsfig{file=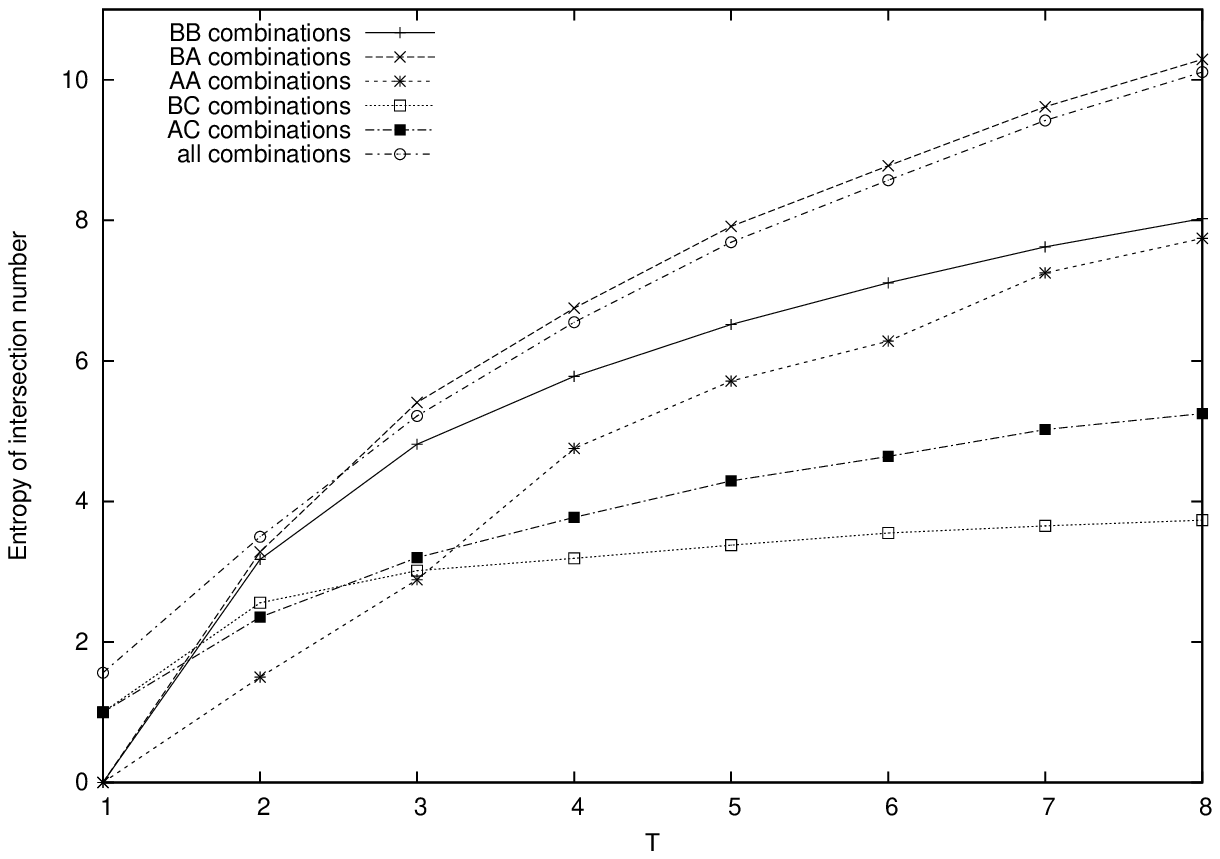,width=15cm}
\caption{\footnotesize Entropies in intersection numbers for different
  brane combinations}
\label{f:2-entropy}
} 
The entropies grow roughly as the log base 2 of the standard
deviations, indicating a reasonably uniform distribution.

While we have computed the intersection numbers for generic $T$
assuming that we are combining single branes, we can also interpret
these results in terms of intersection numbers when the size $N$ of
the brane stacks is increased.  For example, in
Figures~\ref{f:2-sd},~\ref{f:2-entropy} the intersection numbers for a
given value of $T$ are equal to those for the desired physical tadpole
value $T_0 = 8$ and brane stacks of size $N = 8/T$.  For example, for
single \Ab, \Bb\ branes with $T = 4$ the intersection number has a standard
deviation of $\sim 50$.  This is the same as for a pair of brane stacks each
with $N = 2$ with $T = 8$.  The consequence of this is that for larger
gauge groups, the intersection number is polynomially suppressed.
While for a $U(1) \times U(1)$ theory, typical intersection numbers
are in the hundreds (assuming hidden A branes are not included;
including such branes would leave to even larger intersection
numbers), when even a gauge group like $U(2)\times U(2)$ is included
the intersection numbers are reduced by a numerical factor of order
$2^5 = 32$ (the factor in the data graphed in Figure~\ref{f:2-sd} is
slightly smaller than this, presumably due to log suppression of the
rate of growth).  When the gauge group has several factors of
different rank, it is difficult to give a precise estimate for the
suppression of intersection numbers, but for generic configurations
including the possibility of hidden sector A branes, we expect an
overall suppression by a monomial of degree around 7 in the sizes of
the gauge group components.  Thus, typical intersection numbers of a
generic model with for example $U( 4) \times U(2) \times U(2)$ will be
down by a factor of at least $2^7$ from intersection numbers in
generic $U(1) \times U(1) \times U(1)$ configurations.

The information contained in the standard deviation and entropy of the
intersection number distributions give a rough characterization of the
overall structure of these distributions.  
\FIGURE{
\epsfig{file=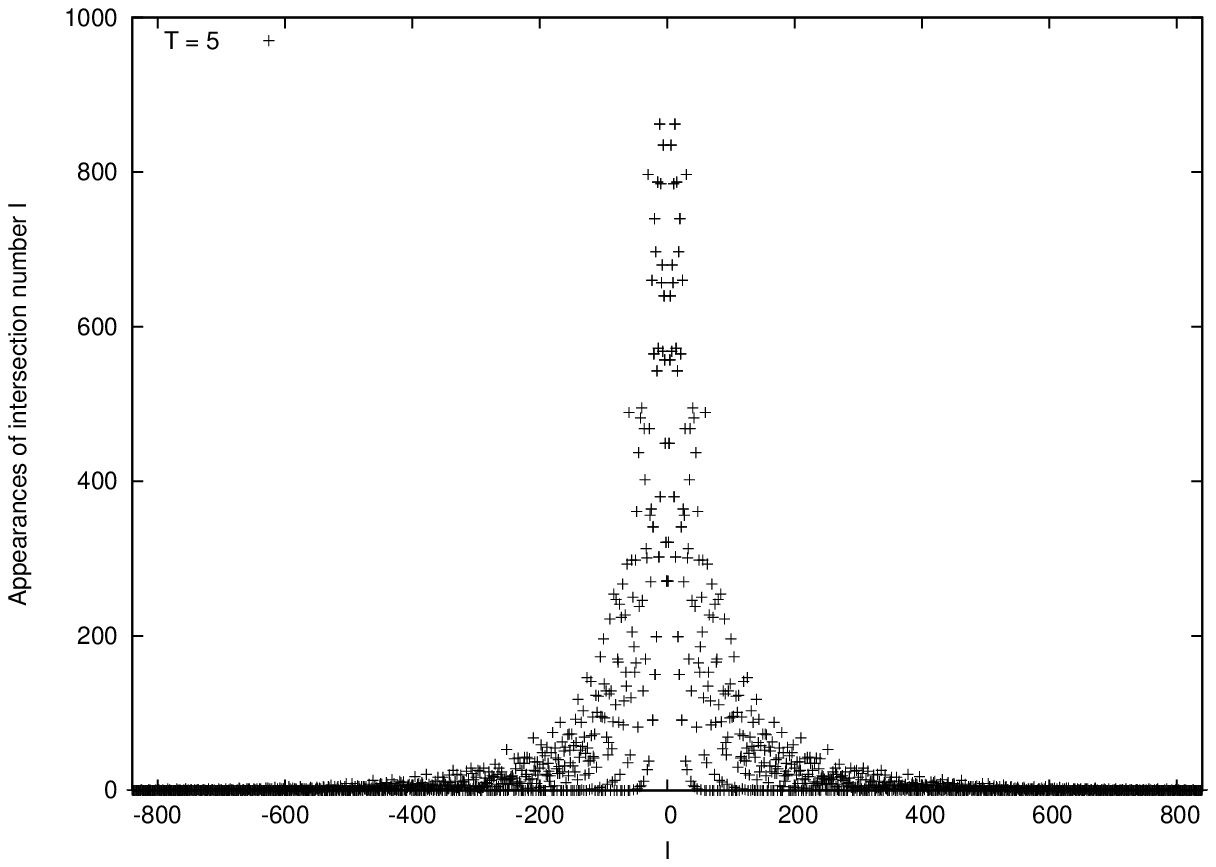,width=15cm}
\caption{\footnotesize Frequencies of intersection numbers at $T = 5$}
\label{f:gi1}
} 
A plot of the overall distribution of intersection numbers for pairs
of branes at $T = 5$ is shown in Figure~\ref{f:gi1}
As expected from Figure~\ref{f:2-sd} the width of the distribution
is of order $100$.  The distribution is fairly sharply peaked, and has
some fine structure.  A closer look at the detailed distribution of
intersection numbers reveals some further features.  In
Figure~\ref{f:gi2} we plot the frequency with which individual
intersection numbers take integer values from 0 to 30 at $T = 5$ and
$T = 6$.  \FIGURE{ \epsfig{file=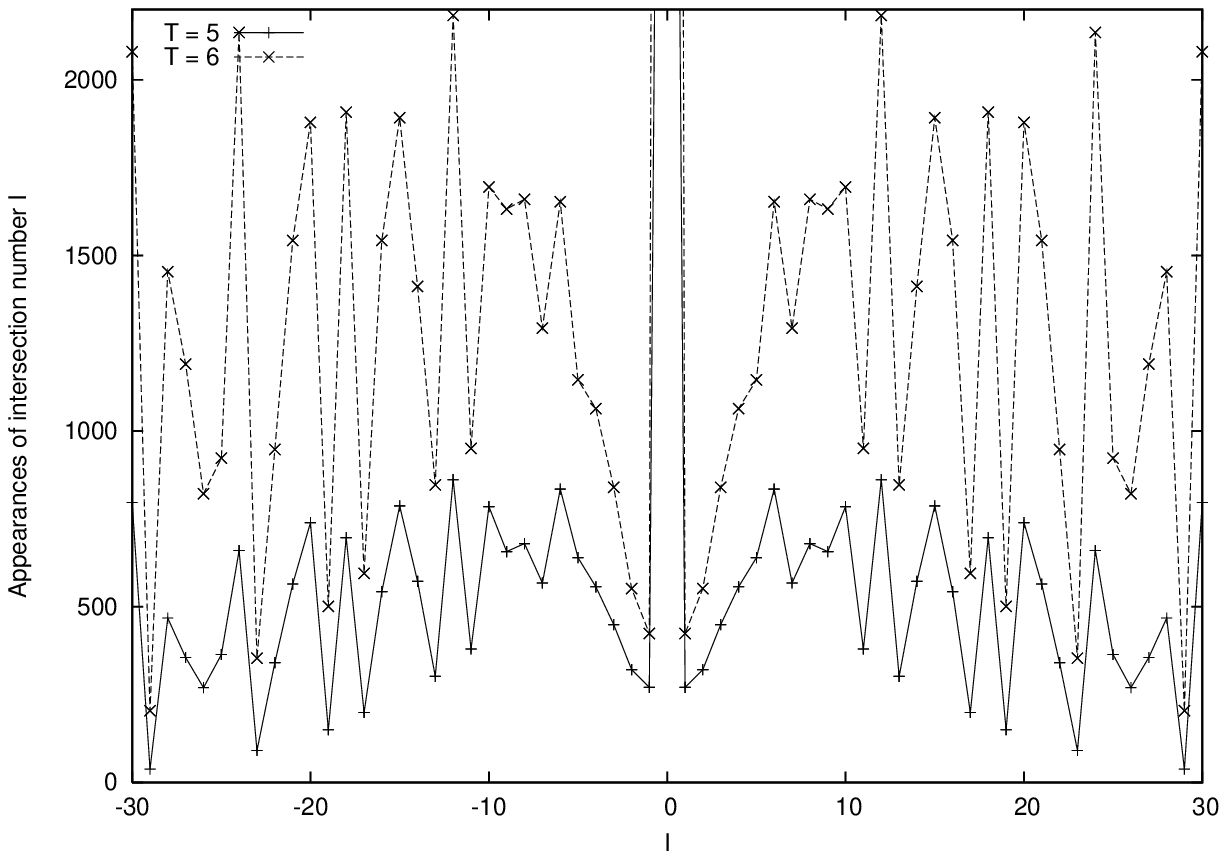,width=15cm}
\caption{\footnotesize Frequencies of small intersection numbers}
\label{f:gi2}
} The pattern shown by this plot is that composite intersection
numbers with many factors are favored, and 0 is very strongly favored,
while prime intersection numbers and other intersection numbers with
few factors are disfavored.  This feature, as well as the sharply
peaked nature of the distribution, comes from the structure of the
intersection number formula (\ref{eq:intersection}), which is a
product of 3 terms each of the form $n \hat{m}-m \hat{n}$.  While this
structure is specific to the toroidal model we are considering here
and presumably is not relevant for more general Calabi-Yau manifolds
with more general intersection formulae, we can briefly give a
quantitative explanation for the relative prevalence of different
small intersection numbers.  Consider for example the parity of the
intersection number, associated with whether it is divisible by 2 or
not.  If the individual winding numbers were chosen with arbitrary
parities, of the 16 possible values for the individual parities, $n
\hat{m}-m \hat{n}$ has parity 0 for 10 combinations and parity 1 for 6
combinations.  This would suggest that only $(6/16)^3 \approx 5\%$ of
the intersection numbers should be odd.  Because the winding numbers
$n_i, m_i$ are relatively prime and not evenly distributed (for
example, 1 is particularly frequent), the true fraction is
significantly larger.  In fact, we have $30\%$ odd intersection
numbers at $T = 4$ and $33\%$ odd intersection numbers at $T = 6$.
Similarly, we generically expect intersection numbers to be congruent
to 0 mod $p$ for each $p$ with a an enhancement factor which decreases
with growing $p$. This number theoretic suppression of intersection
numbers with few factors has a noticeable but relatively mild effect
on the prevalence of different intersection numbers, although it does
lead to a significant enhancement of the probability for a vanishing
intersection number.  Among small intersection numbers it does
suppress 1 more than other numbers, but considering for example the
frequency of appearance of the intersection number 3 across all brane
pairs, we find at $T = 6$ that the intersection number 3 appears about
$0.66\%$ of the time, which is not particularly frequent or infrequent
compared to other numbers in the range from $-100$ to $100$, and which
is quite compatible with the entropy $H \approx 8.57$ for intersection
numbers of pairs of branes at $T = 6$, which suggests that typical
intersection numbers will appear with frequency of order $0.5\%$.  For
comparison, the most common nonzero intersection number $12$, with many
factors, occurs about $1.8\%$ of the time, while a more highly
suppressed larger prime intersection number like $23$ occurs about
$0.27\%$ of the time.  Again, the greatest enhancement is for
vanishing intersection number, which occurs $7.5\%$ of the time.

It is also worth noting that in model building, when we consider two
branes $x$ and $y$ which are not invariant under $\Omega$, the number
of chiral fermions will go as the sum $I_{xy} + I_{xy'}$.  This
addition of intersection numbers tends in general to ameliorate
somewhat the number theoretic suppression or enhancement of certain
values.  The story is different, however, for $p = 2$.  In this case,
we have
\begin{equation}
I_{xy} \equiv I_{xy'} ({\rm mod} \; 2) \;
\Rightarrow \; I_{xy} + I_{xy'} \equiv 0 ({\rm mod} \; 2) \,.
\label{eq:parity}
\end{equation}
As has been discussed in many places \cite{Blumenhagen:2005mu}, 
this implies that
we cannot build a 3-generation model in this orientifold with type 
\Ab\ and \Bb-branes alone without including some additional structure like
discrete $B$ fields which can change these parity constraints.

\subsection{Multiple intersection numbers}

Let us now turn to pairs of intersection numbers.  We are interested
in determining whether there is a strong correlation between distinct
intersection numbers in a given brane configuration, or whether the
intersection numbers are roughly independently distributed according
to the single intersection number distribution described in the
previous subsection.  We consider two possible types of correlation,
first between intersection numbers $I_{xy}$ and $I_{xz}$ in a
configuration of 3 branes $x, y, z$, and second
between intersection numbers $I_{xy}$ and $I_{xy'}$ where $y$
and $y'$ are related through the orientifold action $\Omega$.

As an example of the first type of correlation, consider the most
prevalent 3 brane combination \Ab\Bb\Bb\  (in the absence of hidden A
branes).  Let us consider the ensemble of all such brane combinations
$a + b + \hat{b}$.  The distribution of allowed values of $(I_{ab},
I_{a \hat{b}})$ is shown in Figure~\ref{f:gi3} for $T = 3$.
\FIGURE{ \epsfig{file=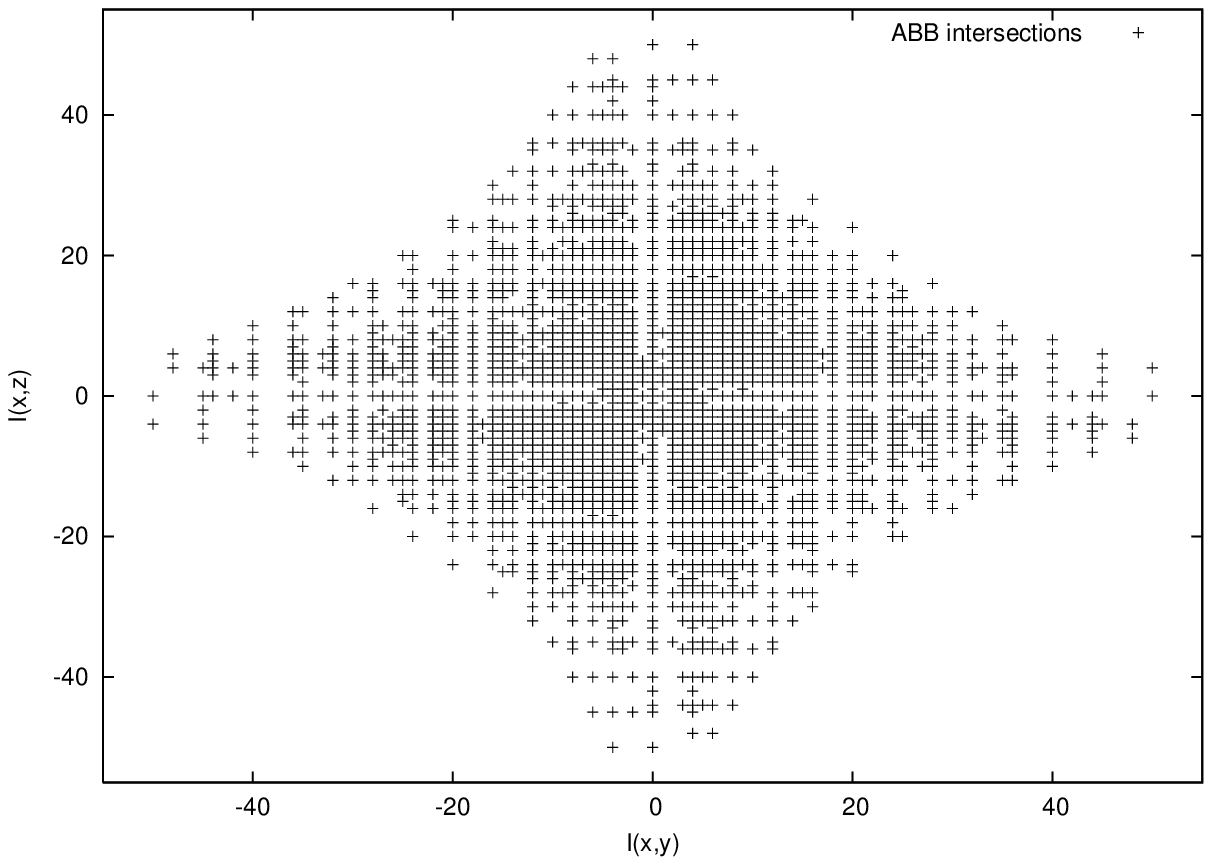,width=15cm}
\caption{\footnotesize Intersection numbers $(I_{xy}, I_{xz})$ for \Ab\Bb\Bb\ 
combinations at $T = 3$}
\label{f:gi3}
}
There are 1594 distinct \Ab\Bb\Bb\  combinations for this value of $T$, and we
consider all pairs of intersection numbers where for each brane we
take either orientifold image.  Including either ordering of the
branes $b, \hat{b}$ gives $16 \times 1594 = 25504$ pairs of
intersection numbers.  We find that the resulting distribution is very
close to a product distribution.  Like the individual intersection
numbers considered above, combinations where either intersection
number is prime are suppressed, and composite intersection numbers
with many factors are enhanced.  To quantify the correlations between
the two intersection numbers, we have computed the mutual entropy
between the distributions of the two intersection numbers.
The distribution of the individual intersection numbers has entropy $H
(I (x, y)) = H (I (x, z)) \approx 4.553$, while the  mutual entropy is just
\begin{equation}
{\cal I} (I (x, y), I (x, z)) \approx 0.085
\end{equation}
Thus, the distributions are
almost completely independent.  We have not done a thoroughly
comprehensive analysis of all brane configurations, but we have looked
at a number of other cases and variations of this calculation, and the
results appear to be fairly general.  For example, for the same
ensemble, the mutual information  between the \Ab\Bb\ and 
\Bb\Bb\ intersection numbers is
${\cal I} (I (x, y), I (y, z)) \approx 0.102$ while the entropy of the $BB$
intersection numbers is $H (I (y, z)) \approx
4.401$.  As we increase $T$,
the ratio of mutual information to total information decreases
further.

We have also looked at the correlation between intersection numbers $I
(x, y)$ and $I (x, y')$ where $y, y'$ are related by $\Omega$.  In
this case the correlations are somewhat stronger.  From
(\ref{eq:parity}), the parities of the these intersection numbers must
be the same.   There are also further correlations arising from the
fact that the winding numbers of $y$ and $y'$ are the same up to signs.
For the class of \Ab\Bb\ configurations, which dominate two-brane
combinations in the absence of hidden \Ab\ branes, we find that at $T =
4$, the entropy of each intersection number is $H (I (x, y)) \approx
6.6$, and the mutual entropy is ${\cal I} (I (x, y), I (x, y'))
\approx 3.4$, so knowing one intersection number gives almost half the
information needed to determine the other.  The ratio of mutual information
to total entropy decreases as $T$ increases; at $T =6$ we have $H
\approx 8.7$ and ${\cal I} \approx 4.0$.

Finally, we have looked at some sets of 3 intersection numbers to see
if correlations between three intersections are stronger than those
between two intersection numbers.  In particular, we consider again
the class of \Ab\Bb\Bb\ configurations without hidden sector \Ab\ branes.  
In principle, we could compute the 3-way mutual entropy for the three
intersection numbers.  This would require a substantial amount of data
for a useful test, so we have done a simpler spot test.
We can select all configurations with a fixed intersection number, say $I
(y, z)= I_0$ and ask whether the distribution of intersection numbers
$I (x, y), I (x, z)$ is significantly different from that without the
constraint on $I (y, z)$.  We did this with all 49,106 \Ab\Bb\Bb\ %
combinations at $T = 4$ for the values $I_0 = 2$ and $I_0 = 3$, which
correspond respectively to $2504$ and $4360$ combinations.  In each case we
found that the distribution of intersection numbers $I (x, y), I (x,
z)$ was qualitatively similar to the distribution without constraining
$I (y, z)$.  For example, at $I_0 = 3$, the mutual entropy was about
0.4, with individual entropies for the \Ab\Bb\ intersection numbers of
about 6.3.  This suggests that even when 3-way correlations are
considered there is very little connection between different
intersection numbers.

The upshot of the analysis described in this section is that for all
the brane combinations we considered here, the intersection numbers
were fairly independent.  For intersection numbers $I (x, y), I (x, z)$
between a fixed
brane $x$ and two other branes $y, z$ which are not related by the orientifold
symmetry and which live in  an allowed combination, like the \Ab\Bb\Bb\ %
combinations, the intersection numbers are almost completely independent.
We found somewhat stronger correlations between intersection numbers
$I (x, y), I (x, y')$ where $y$ and $y'$ are related by the
orientifold action.  These correlations include the fact that such
pairs of intersection numbers must have the same parity, giving an
even number of fermions living in this sector of the theory.

The basic conclusion which follows from this analysis is that if we
wish to estimate the number of models which have a given gauge group
and given numbers of chiral fermions in each matter sector
corresponding to pairs of intersecting branes, we can get a good
estimate by multiplying the total number of models with the desired
gauge group by the product of the fractions of those models with the
desired number of fermions in each sector.  This provides a rough
estimate of the number of models with the desired properties which can
be constructed using intersecting branes in this class of
orientifolds.  Furthermore, we can use the algorithms described in the
previous section to explicitly enumerate the desired intersecting
brane configurations.  By imposing the intersection number conditions
as we include branes, we can produce an algorithm which enumerates all
desired brane configurations with given intersection numbers in a time
which scales roughly as the number of solutions, with possibly a small
polynomial overhead.  

\section{Some specific models}
\label{sec:models}

We now have estimates for the number of models which can be
constructed containing a particular gauge group, and we have analyzed
the general structure of the distributions of intersection numbers.
We have an upper bound for the number of models possible with a fixed
gauge group such as $SU(3) \times SU(2) \times U(1)$ or $SU(4) \times
SU(2) \times SU(2)$ which is of order ${\cal O} (T^{13}/N_1^2 N_2^2
N_3^9)$, with $N_3$ the size of the smallest factor and $T = 8$.
Thus, we expect at most order $10^{10}$ or so configurations
containing the gauge group $SU(3) \times SU(2) \times U(1)$, and order
$10^7$ configurations containing the gauge group $SU(4) \times SU(2)
\times SU(2)$.  The precise numbers of these configurations can
computed efficiently in a time of order of the number of
configurations using the algorithms discussed in Section
\ref{sec:algorithms}, and the resulting configurations can be
tabulated.  From the analysis of Section \ref{sec:intersection}, we
know that the intersection numbers are roughly independently
distributed, with each intersection number having values typically of
order $(T/N)^7$, and with some mild suppression of prime intersection
numbers and enhancement of intersection numbers with many factors.
Fixing 3 independent intersection numbers to take values of order
$1-10$, then, in a model with one of the above gauge groups, should
reduce the number of models by a factor of something like $10^9$ for
$SU(3) \times SU(2) \times U(1)$ (using $N = 3$) or $10^6$ for $SU(4)
\times SU(2) \times SU(2)$ (using $N = 4$).  In either case, this
leads us to expect something like order 10 models with the desired
gauge group and intersection numbers.

A natural next step in this analysis might be to perform a systematic
search using the methods we have described here for models with the
standard model or Pati-Salam gauge group and 3 generations of chiral
leptons and quarks in the appropriate representations of the gauge
group.  To begin to construct semi-realistic models, however, we must
take into account the parity constraint (\ref{eq:parity}), which
states that $I_{xy} + I_{xy'} \in 2\iz$ if $y$ is a type \Ab\  or type 
\Bb\ brane.  Since the number of chiral fermions in the $U(N_x) \times
\bar{U} (N_y)$ representation is given by this sum, we see that the
parity constraint makes it impossible to have an odd number of
generations when $y$ is type \Ab\  or \Bb.  Thus, we cannot immediately
construct the expected handful of standard model-like brane
configurations as discussed above, since their generations will all be
even.

There are two ways in which we can evade the parity constraint and
generate odd generation numbers.  The first is to include discrete $B$
fields on some of the two-tori forming the $T^6$ (in the T-dual
picture this corresponds to skewing the tori). 
This
approach was taken in \cite{Cvetic:2001nr,Cveticll}, where they systematically
searched for constructions with the standard model or Pati-Salam gauge
group and 3 generations of quarks and leptons.  Because the inclusion
of discrete $B$ fields only slightly changes the equations, basically
shifting the winding modes in the affected tori by $m_i \rightarrow
m_i + n_i/2$, we expect that the qualitative features of the analysis
we have carried out in this paper go through unchanged, so that in the
presence of discrete $B$ fields, while the parity constraint no longer
rules out 3 generations, the expected number of models is still of
order ${\cal O} (10)$.  This is in nice correspondence with the
results of \cite{Cveticll}, where 11 such distinct models were
identified.  

The second way of including odd generation numbers is to put part of
the standard model or Pati-Salam group on \Cb-type branes, which do
not have extra image branes, and therefore are able to have odd
fermion numbers.  This approach was taken in
\cite{Cremades:2003qj,Marchesano:2004yq,Marchesano-Shiu-2,Cvetic:2004nk}
where they constructed a $U(4) \times U(2) \times U(2)$ Pati-Salam
model which in our language is constructed with a $U(4)$ coming from a
stack of 4 \Bb\ type branes, and two $U(2)$ factors coming from \Cb\
type branes.  The \Bb\ type brane oversaturates a tadpole, but can be
compensated by a hidden sector \Ab-brane.  In \cite{Kumar-Wells}, they
carried out an analysis of other hidden sector \Ab-brane combinations
which could also work with this construction.  A similar construction
of an $U(4) \times U(2) \times U(1) \times U(1)$ model was given in
\cite{Chen:2005mj}.

{}From the general analysis of this paper it is clear that  another way
to realize a 3-generation Pati-Salam model is to take the $U (4)$ to
come from a stack of 4 \Ab-type branes.  
Let us assume that we have two factors of $U(2)$ coming from \Cb-type
branes $c$ and $\hat{c}$ with tadpoles $R = 1$ and $S = 1$.  From
(\ref{eq:intersection}), the condition that an \Ab-brane $a$ has
intersection numbers $I_{ac} = I_{a \hat{c}}= 3$ implies that
\begin{equation}
-n_1 m_2 n_3 = -n_1 n_2 m_3 = 3 \,.
\label{eq:3-conditions}
\end{equation}
These conditions are satisfied if we take a
stack of $N = 4$ \Ab\ branes with
\begin{equation}
n = (3, 1, -1),  \;\;\;\;\;
m = (1, 1, -1), \;\;\;\;\;
(P, Q, R, S) = (-3, 3, 1, 1) \,.
\label{eq:4-as}
\end{equation}
along with their orientifold images with $m \rightarrow -m$.  Other
solutions of (\ref{eq:3-conditions}) are possible but require two
different winding numbers to be 3, leading to more than one tadpole
which is oversaturated.  Taking the stack of 4 \Ab-branes
(\ref{eq:4-as}) and their images and the two pairs of \Cb\ type branes,
we have total contributions to the tadpole $(-12, 12, 5, 5)$.  We need
to include at least one additional ``hidden sector'' \Ab-type brane to
cancel the excess tadpole in $Q$.  We have systematically searched for
such hidden branes, and found 17 possibilities for this hidden sector
brane, which are listed in Appendix  B.  There are presumably many
other possible hidden sector combinations of multiple \Ab\ branes, but we
have not searched for those.

A further check which must be performed to confirm that the models we
have constructed with a single hidden sector \Ab-brane are allowed is to
confirm that these brane combinations either satisfy the K-theory
constraints (\ref{eq:K-theory})
or satisfy these constraints after a further set of branes is
included.  Once we have a set of branes which satisfy the K-theory
constraints and undersaturate all tadpoles, the brane combination can
be extended to one which precisely cancels all tadpoles by adding type
\Cb\ branes, which do not contribute to the K-theory constraints
(\ref{eq:K-theory}).  None of the 17 possible combinations listed in
Appendix 2 satisfy the K-theory constraints without the addition of
further branes.  We have checked all 17 models for combinations which
satisfy the K-theory constraints as well as the SUSY conditions with
the addition of a single type \Bb\ brane.  Of the 17 models, only the 5th
can be combined with a single type \Bb\ brane  to satisfy all the
K-theory constraints.  In this model, we start with a stack of 4 \Ab-type
branes as in (\ref{eq:4-as}), \Cb-branes with $R = 1$ and $S = 1$, and
a single \Ab-type brane with
\begin{equation}
n = (2, 1,3),  \;\;\;\;\;
m = (1, -1 -2), \;\;\;\;\;
(P, Q, R, S) = (6, -4, 2, 3) \,.
\label{eq:first-a}
\end{equation}
So far this gives total tadpoles $(-6, 8, 7, 8)$.  We can then add
either a single B brane with
\begin{equation}
n = (14, 1, 1),  \;\;\;\;\;
m = (-1, 0,1), \;\;\;\;\;
(P, Q, R, S) = (14, 0, 1, 0) \,
\end{equation}
which fixes all the K-theory constraints and precisely saturates all
tadpoles, or we can add a single B brane with
\begin{equation}
n = (12, 1, 1),  \;\;\;\;\;
m = (-1, 0,1), \;\;\;\;\;
(P, Q, R, S) = (12, 0, 1, 0) \,
\end{equation}
which fixes all the K-theory constraints and requires an additional
pair of (really 4) type \Cb\ branes with $P = 1$ to cancel all tadpoles.

The class of models we have constructed here are not completely
realistic, in that they contain chiral exotics which are difficult to
remove from the low-energy spectrum (see, {\it e.g.}
\cite{Marchesano-Shiu-2, Kokorelis:2004tb}).  Nonetheless, they
provide an illustration of how the methods we have developed can be
used to construct quasi-realistic models with certain desired
features.  The construction described here will presumably lead to
some other 3-generation Pati-Salam models if more than one hidden
sector \Ab-brane or more than one hidden sector \Bb\ brane is allowed,
but we have not investigated this possibility further.  We content
ourselves for now with this pair of new possibilities as examples of
how the techniques developed in this paper allow us to systematically
search for intersecting brane models with given gauge group and
numbers of chiral fermions.  Like other such constructions, the models
we have constructed have various ``hidden sector'' gauge groups and
chiral matter fields transforming under the hidden gauge groups.  In
the UV, anomalies are cancelled by the Green-Schwarz mechanism, but in
the low-energy field theory, generally these models will be anomalous,
and much of the hidden sector will be lifted to a high energy scale,
leaving a residual anomaly-free low-energy field theory.  We leave a
detailed analysis of the models we have described here for future
work, our interest here is primarily as examples of the methodology
developed herein and not on detailed phenomenology.

\section{More general Calabi-Yau orientifolds}
\label{sec:notation}

The type of analysis we just discussed can be directly generalized to
orientifolds of other Calabi-Yau manifolds, given a sufficiently large
set of supersymmetric branes.  While at this point it is not clear how
large a set would suffice to get a representative sample, probably the
minimal requirement is to have a family with at least $b^2$ parameters
(in IIB terms), so that the ratio between the tadpole contributions
can be adjusted independently (their overall scale can be adjusted by
replicating the branes).  It would be nice to have a family of branes
with all $2b_2+2$ homology classes freely adjustable.  Note that for
$T^6/\IZ_2^2$ (leaving out fractional branes) we had $6$ winding
numbers while $2+2b_2=8$, so we did not quite meet this criterion even
there.

In the IIA picture, the supersymmetric branes are D6-branes
wrapping special Lagrangian submanifolds.  The only general
construction we know for these is to embed the brane in the fixed
point of an antiholomorphic involution, which is not sufficiently general.

In the IIB picture, the supersymmetric branes wrap holomorphic
submanifolds of the CY, and carry holomorphic bundles.  Now there is a
simple class of bundles we can construct on any smooth CY, namely the
holomorphic line bundles.  These were first suggested for model
building in \cite{Bachas} and have been much used since.  On a simply
connected CY threefold, a line bundle is parameterized by its first
Chern class, an element of $H^2(M,\IZ)$.  Thus the line bundles
provide a family of branes with $b_2$ parameters.  

As we mentioned in subsection \ref{sec:SUSY-branes}, on $T^6$ the line
bundles are mirror to D$6$-branes with all winding numbers $n_i=1$,
so this is a natural subset of the brane configurations we
considered earlier.  One might ask why these simple bundles were not
used in the original heterotic string constructions.  The answer is
that in the weak coupling and large volume limit, the supersymmetry
condition Eq. (\ref{eq:basic-SUSY}) does not admit interesting
solutions.  We saw this implicitly in the IIB analysis above, in that
solving the full conditions require finite K\"ahler moduli.  It is
also true in the heterotic string, as the $\mu$-stability condition on
bundles is exactly what one obtains by dropping the $m_1m_2m_3$ term
from Eq. (\ref{eq:basic-SUSY}), and thus a $\mu$-stable direct sum of
line bundles would give a (non-existent) IIB brane solution in the
large volume limit.

Recently in \cite{Blumenhagen:2005ga}, it was shown that heterotic
string loop effects produce a correction to the $\mu$-stability
condition which allows sums of line bundles to provide supersymmetric
solutions there as well.  In the $SO(32)$ string this is of the same
form as Eq. (\ref{eq:basic-SUSY}) (indeed it is S-dual to it), while
in the $E_8\times E_8$ string it is similar.  We mention this in passing
as a piece of evidence that all of the various known constructions
are just parts of a larger (and one hopes simpler) classification.

Another natural family is the D$5$-branes wrapping holomorphic curves.
The homology class of a curve also has $b_2$ parameters; of course there
can be many curves in a given class so there are further discrete 
choices.\footnote{In orientifold theories besides type I, it may also
be possible to turn on magnetic flux $F$ on the D$5$.}

In the following, we will discuss the problem of enumerating sums of
line bundles and curves satisfying the tadpole cancellation conditions
on general Calabi-Yau orientifolds.  Let us briefly mention two other
interesting sets of branes.  One is the rational boundary states in
Gepner models; these have been used in \cite{Schellekens}.
These correspond to certain bundles obtained by the
generalized McKay correspondence for the corresponding orbifold of
$\IC^5$, in a way which is discussed in 
\cite{Diaconescu:2000ec}; the generalization of this to orientifolds is
partially understood \cite{Brunner:2004zd}.

A second interesting class of branes, which makes sense for a
Calabi-Yau manifold which is a subvariety of a toric variety, is the
restriction of the equivariant bundles on the toric variety.  This
includes the large set of toric hypersurfaces, complete intersections
in toric varieties, and so on, so it applies very generally.  In
words, we are using the fact that such a Calabi-Yau manifold admits a
simple embedding into a higher dimensional space (the toric variety)
with a large symmetry group (a product of $U(1)$'s); the equivariant
bundles are then the bundles on which this symmetry group acts by
gauge transformations.  This includes the line bundles, and loosely
speaking, the generalization corresponds in the case of $T^6$ to
lifting the restriction on the winding numbers $n_i=1$, to allow
general $n_i$.  Having said this, however, we leave further discussion
of this class to future work.

\subsection{General framework}

We will now set up a general framework for enumerating type IIB
orientifold compactifications using any set of bundles on any
Calabi-Yau compactification.  Another description of this
class of compactifications can be found in \cite{Blumenhagen:2005zh}.
We will assume various standard
mathematical definitions and be somewhat telegraphic; it may be
helpful to look ahead at the sections on the torus and on the
Calabi-Yau examples.

\subsubsection{Basic definitions}

\begin{itemize}

\item $M$ is the the Calabi-Yau threefold.

\item $\CJ\cong H^{1,1}(M,\IR)$, and $J\in \CJ$
is a K\"ahler form.  In type II theory this would be complexified
by the $B$ field, but the orientifold projection removes this
continuous degree of freedom.  However, there is still a discrete
choice of $B=0$ or $B=1/2$ on each two-cycle consistent with this projection.
This can summarized in an element $2B\in H^2(M,\IZ_2)$.

\item $F\in\CF\equiv\CJ \cap H^2(M,\IZ)$ is a
normalized magnetic field strength, $F=dA$ for a $U(1)$ connection.

\item By Poincar\'e duality, $H^2(M,\IZ)\cong H_4(M,\IZ)$, so we can also think
of this datum as the homology class of a two complex dimensional submanifold
$D$ of $M$.  
A line bundle with magnetic field $F$ will be denoted either
$\CO(D)$ or $\CO(F)$.  
Such a submanifold is called a ``divisor'' in algebraic geometry; a generic
section of the line bundle $\CO(D)$ will vanish on $D$.
Physically, one can think of the D$9$-brane carrying such a
line bundle, as a bound state of the ``elementary'' (zero flux) D$9$ with
a D$7$ wrapped on $D$.

\item $\{\omega^i\}$ is an integral basis for $\CF$, so that a
field strength can be written as $F=\sum_i f_i \omega^i$.  One can also
introduce an integral basis of divisors $D_i$.  

\item The dual bundle $V^\vee$ to a bundle $V$ is defined by inverting
the transition functions.  For a line bundle $\CO(F)$,
we have $\CO(F)^\vee=\CO(-F)$.  

\item If $2B$ as defined above is not zero, the field strength $F$ which
appears in all of the following definitions should be replaced by 
$F-B\in H^(M,\IR)$.  This can be defined by choosing a lift of $2B$ to
$H^2(M,\IZ)$ and taking $(2F-2B)/2$.

\item $\CK\equiv \oplus_p H^{p,p}(M,\IR)\cap H^{2p}(M,\IZ)$
is the lattice of even homology classes.  More precisely, this
should be $K^0(M)$, but this matters only for the torsion,
so we ignore it for now.  

\item $\Omega$ is the orientifold operation on a bundle.
In type I theory, this is the same as the dual, thus
$\Omega \CO(F)=\Omega(-F)$.  In other IIB models, $\Omega$ acts
as a reflection on $\CK$.

\item We denote the Chern character (K theory class) of a bundle 
$E$ as $[E] \in \CK$.  It is additive under direct sum.
For a vector bundle $E$ with curvature $F$,
$$
[E]=\tr e^F.
$$
For the special case of a line bundle $\CO(F)$ this is just $e^F$.

\item The K theoretic intersection product on $\CK$ is
$$
\eta(K,L) = \int_M K^\vee\cdot L\cdot \hat A(M)
$$
where $\hat A(M)$ is the A-roof genus as in the index formula,
$\hat A(M) = 1 + \frac{1}{24}c_2(M)$.
It is antisymmetric and integral on $\CF^2$, and determines the
number of bifundamental multiplets between branes carrying the
bundles $K$ and $L$.

\end{itemize}

\subsubsection{Calabi-Yau geometric data}

So far, we have just been giving names to various standard
spaces and constructions; the only dependence on $M$ is through
its second Betti number $b_2=\dim H^2(M,\IR)$.

The non-trivial geometric data we need to know about $M$ to construct brane
theories is the following:
\begin{itemize}

\item $C(\alpha,\beta,\gamma): \CJ^3\rightarrow\IR$
is the cubic intersection form, defined as
$$
C(\alpha,\beta,\gamma) = \int_M \alpha\wedge\beta\wedge\gamma.
$$
It takes integer values on $\CF^3$.
Unless otherwise denoted, products on $\CJ$, $\CF$ and so on are in
terms of this form.  

One can equivalently think of this in Poincar\'e dual terms
as $C(D_i,D_j,D_k)$,
the number of points in which three divisors intersect.
{}From this point of view, one can see that $C(D_i,D_j,D_k)\ge 0$ for
three distinct holomorphic submanifolds.

\item The second Chern class $c_2(M) \in H^4(M,\IZ)$.  This is
Poincar\'e dual to a class in $H_2(M,\IZ)$, and an equivalent way
to give this information is the set of intersection numbers
$c_2(M)\cdot D_i$ for a basis of divisors.

\item The K\"ahler cone $KC(M)\subset H^2(M,\IR)$ is the set of
K\"ahler forms of sensible K\"ahler metrics (all integrals $\int \omega^k$
over holomorphic cycles are positive).  It is convex, {\it i.e.}
given any $\omega_1,\omega_2\in KC(M)$, we have 
$\omega_1+\omega_2 \in KC(M)$.

\item The Mori cone $MC(M)\subset H_2(M,\IZ)$ given by the 
effective classes; {\it i.e.} the classes of holomorphic curves in $M$.
By Poincar\'e duality we can identify it with a subset of $H^4(M,\IZ)$.
It is also a convex cone.

It turns out that the K\"ahler cone and the Mori cone are dual, in the
sense that $\omega \in H^2(M,\IR)$ is in the K\"ahler cone if
$$\eta(\omega,C) > 0 \qquad \forall C\in MC.$$
Thus either can be deduced from the other.

\end{itemize}

This data is used as follows.  We choose a set of branes $B_\alpha$, and
consider a configuration
$$
B = \sum_\alpha N_\alpha B_\alpha
$$
where $B_\alpha$ are distinct branes and $N_\alpha$ are multiplicities.
In type II theory this configuration would have gauge group 
$\prod_\alpha U(N_\alpha)$,
before taking into account anomalous $U(1)$'s.

In a type I model, the brane configuration must satisfy
$$
\Omega B = B^\vee = B ,
$$
and tadpole cancellation,
\begin{equation}\label{eq:gen-tadpole}
[B] \equiv T = 16 + c_2(M) ,
\end{equation}
where ``$1$'' is the generator of $H^0(M,\IZ)$.

This is an equation between elements of $\CK$, whose expansion into
components is the set of tadpole constraints.  To make this a bit more
concrete, suppose that each of the branes $B_\alpha$ is a D$9$-brane
carrying a line bundle with divisor $D_a=\sum_i m^i_a D_i$, and comes with
an image $B'_\alpha$ with divisor $-D_a$.  Then the tadpole conditions
become
$$
8 = \sum_\alpha N_\alpha
$$
and
$$
T_i \equiv  c_2(M) \cdot D_i =
 \sum_a N_a C(D_a,D_a,D_i) 
$$
where $D_i$ runs over a basis of divisors.

For a supersymmetric compactification
in the large volume limit, each brane must satisfy
$$
Z_\alpha \equiv C(F_\alpha+iJ,F_\alpha+iJ,F_\alpha+iJ)
 = -i\lambda_\alpha \qquad \lambda_\alpha\in\IR^+
$$
where $J\in H^2(M,\IR)$ is the K\"ahler class.  
The central charge $Z$ could equivalently be written as
$$
Z = \eta(e^{-iJ},[B_i]) .
$$

For reasons we discuss momentarily, here we restrict attention to
the large volume limit, in which these conditions can be broken into
\begin{equation}\label{eq:re-gen-susy}
0 = \Re Z_\alpha = -3 C(J,J,F_\alpha) + C(F_\alpha,F_\alpha,F_\alpha)
\end{equation}
and
\begin{equation}\label{eq:im-gen-susy}
0 < -\Im Z_\alpha = C(J,J,J) - 3 C(J,F_\alpha,F_\alpha).
\end{equation}

Unlike the torus, where these equations are exact, on a general
Calabi-Yau they will receive world-sheet instanton corrections.  By
the ``large volume limit'' we mean the limit in which these can be
dropped, which is when the volumes of all holomorphic curves are large
in string units.  If one goes away from the large volume limit, there
are further subtleties in the discussion because supersymmetric branes
are in general not branes carrying bundles but instead $\Pi$-stable
objects in the derived category (see \cite{Aspinwall,Clay} for a
discussion of this).  Since many if not most of the models obtained by
the construction discussed here will stabilize K\"ahler moduli at the
string scale, these subtleties will generally be important.

While a detailed discussion is not easy and depends on the example, a
subset of models for which the large volume equations might give a
reasonable description is to only allow D$9$-branes for which the
supersymmetry conditions admit simultaneous solutions at large volume,
as well as D$5$-branes, while forbidding D$7$-branes,.  This is because
D$7$-branes cannot be
simultaneously supersymmetric with the others
without large $\alpha'$ corrections, and furthermore they
are known to decay into the others
at string scales (again, see \cite{Aspinwall}
for examples and references).  Admittedly this is a hypothesis which
would require some justification in examples before accepting, but
based on the current knowledge we think it is reasonable to explore.
Of course one can also deal with this point by restricting attention
to configurations which stabilize the K\"ahler moduli at large volume.

The generalization of \eq{gen-tadpole} to include D5-branes wrapping
holomorphic cycles is simple; now $[B]$ is simply the class of $B$ in
$H_2(M,\IZ)$.  Furthermore, at least in the large volume limit, the
D$5$ branes in type I theory are guaranteed to respect the same
supersymmetry as the O$9$-plane, independent of the K\"ahler moduli.
However, unlike the torus, this statement can get 
corrections from world-sheet instantons.  An interesting concrete example
using these in which the subtleties we mentioned are addressed is  
\cite{Diaconescu}, in which
these corrections are used to break supersymmetry.

\subsubsection{Matter content}

Given a consistent configuration, we can study the matter
content.
Open strings stretched between branes give rise to matter in a
variety of two-index tensor representations constructed from the
fundamental $E_i$ and antifundamental $\bar E_i$.
According to the index theorem, the total matter content is
\begin{eqnarray*}
 E_{ij}\equiv E_i\otimes \bar E_j & \qquad & \eta(B_i,B_j) \\
 P_{ij}\equiv E_i\otimes E_j & \qquad & \eta(B_i,\Omega B_j) \\
 S_i\equiv {\rm Sym}^2 E_i & \qquad &
  \half\left( \eta(B_i,\Omega B_i)- \eta([\Omega],B_i) \right) \\
 A_i\equiv \wedge^2 E_i & \qquad & 
  \half\left( \eta(B_i,\Omega B_i)+ \eta([\Omega],B_i) \right) 
\end{eqnarray*}
where $[\Omega]=1+...$ is the class of the orientifold plane.

This spectrum can be usefully summarized in the generalized
intersection matrix
$$
\hat I(B_i,B_j) = \eta(B_i,B_j) + \eta(B_i,\Omega B_j) .
$$
The two terms are antisymmetric and symmetric matrices
respectively, so this does not lose information.

\subsection{Examples}

\subsubsection{Torus}

Explicit formulas for the $\IZ_2^2$ invariant subsector of $T^6$:
\begin{itemize}

\item $\CJ\cong \IR^3$ with integral basis 
$\omega_i\equiv dz^i\wedge d\bar z^i/2i\Im\tau_i$.
The intersection form is the totally symmetric form with
$$
C(\omega_1,\omega_2,\omega_3)=1
$$
and all other entries zero.  Equivalently, $c_{123}\equiv 1$ etc.

\item $\CK\cong \IZ^8$, and is the algebra 
over the integers generated by three {\it commuting} variables
$\omega_1,\omega_2,\omega_3$ each satisfying $\omega_i^2=0$.
Defining
$\omega_i^\vee=-\omega_i$ and $\int a$ as the coefficient of
$\omega_1\omega_2\omega_3$, we have
$$
\eta(K,L) = \int K^\vee\cdot L.
$$
In components, we write a vector
$$
K=(r,\vec f,\vec m,c)
$$
and the intersection form is then
$$
\eta(K,K') = r c' - c r' + \vec f\cdot \vec m'  -  \vec f'\cdot \vec m .
$$

\item The orientifold operation takes $F\rightarrow -F$ and thus
$$
K=(r,\vec f,\vec m,c) \rightarrow \Omega K=(r,-\vec f,\vec m,-c) .
$$

\item A line bundle is characterized by $F=\sum_i m^i \omega_i$
with $m_i\in\IZ$.  Its Chern character is
$$
[e^F] = (1,m^i,\half c_{ijk}m^j m^k,\frac{1}{6} c_{ijk} m^i m^j m^k)
$$
with integer entries.

\end{itemize}

This corresponds to the subset of the branes we were using earlier
with all $n_i=1$.  Let us see what the rest of the branes look like.
First, the other \Cb-type branes with two $n_i=0$ are clearly D$5$-branes
wrapping the $T^2$ with $n_i=1$.

On $T^2$, there is a unique $U(n)$ bundle $V_{n,m}$ with 
constant field strength and $c_1(V_{n,m})=m$.  It is the T-dual of
a $D1$ brane with winding numbers $(n,m)$.  Thus the T-dual of a
D$6$-brane in IIa theory is the tensor product bundle
$$
V_{\vec n,\vec m} = \otimes_i V_{n_i,m_i} 
$$
with rank $n_1n_2n_3$.  Its Chern character is the tensor product
of the three sets of winding numbers,
$$[V_{\vec n,\vec m}] = 
\otimes_{i=1}^3 \left(\matrix{m_i\cr n_i}\right) .
$$

This accounts for all the \Ab-type branes, and the \Bb-branes with all
$n_i\ne 0$.  The \Bb-branes with one $n_i=0$ are D$7$-branes carrying
bundles in IIb terms.  As we discussed earlier, on more general
string-scale Calabi-Yau manifolds the large volume discussion will
tend to break down for these branes.

One can check that the supersymmetry conditions agree with those we
were using earlier (possibly up to signs in the definitions).  To
check the tadpole conditions, one needs to compute $c_2(M)$ for $M$ a
smooth resolution of $T^6/\IZ_2\times\IZ_2$.  This was done for a
related but slightly different case (compactification with D7 and D3
branes) in \cite{Denef:2005mm}.  There, it was found that matching the
orientifold tadpole implies the presence of O3 planes in the resolved
space as well as O7 planes.  The standard $T^6/\IZ_2\times\IZ_2$
orientifold of our previous discussion has O5 plane tadpole
contributions in the IIB picture which presumably persist in a similar
fashion even after the resolution.  This generalization can be made
for Calabi-Yau as well, but we save this for future work.

The intersection matrix between a pair of these branes is
$$
I(V,V') = \prod_{i=1}^3 (n_i m'_i - m_i n'_i) .
$$
Furthermore,
$$
I(V,\Omega V') = -\prod_{i=1}^3 (n_i m'_i + m_i n'_i) .
$$

The generalized intersection matrix is
$$
\hat I_{ij} = \int [B_i^\vee] \cdot \left([B_j] + [\Omega B_j]\right)
$$
We have
$$\label{eq:genint}
\hat I(K,K') = -2( c r'  +  \vec f'\cdot \vec m )
$$
The overall coefficient $2$ here is correct and thus if $\hat I$
has any odd entries, we are forced to turn on $B$ field.

\subsubsection{Elliptically fibered Calabi-Yau manifolds}

Let us give another simple example\footnote{
We thank Bogdan Florea for providing this example.}
with $b_2=3$, the elliptic
fibration over $F_0 = P^1 \times P^1$.
Let $D_1$ and $D_2$ be the divisors which are the elliptic fibrations over
the two $P^1$'s in the base, and let the third divisor be
$D_0=B+2D_1+2D_2$, where $B$ is
the section of the elliptic fibration. Then, the nonzero
intersection numbers are
$$
D_0^3 = 8,\,\, D_0^2 D_1 = D_0^2 D_2 = 2,\,\, D_0 D_1 D_2 =1.
$$
Taking the K\"ahler class to be
$$
J=\sum_{i=0,1,2} j_i D_i ,
$$
the volume of the Calabi-Yau is
$$
{\rm vol} = \frac{1}{6} J^3 = 
 \frac{4}{3}j_0^3 + j_0^2 j_1 + j_0^2 j_2 +  j_0 j_1 j_2.
$$
For a consistent metric, all holomorphic cycles must have positive volume;
the subset of $j_i\in\IR^3$ in which this is true is called
the K\"ahler cone.  We chose our basis of divisors so that the K\"ahler 
cone turns out to be simply the region with all $(j_0,j_1,j_2)$ positive,
but in general it is defined by a more complicated set of linear inequalities.

The intersections with $c_2(M)$ are given by:
$$
c_2 B = -4,\,\, c_2 D_1 = c_2 D_2 = 24
$$
and thus the tadpole constraints become
\begin{eqnarray}
8 &=& \sum_a N_a \label{eq:fzerotadzero}\\
-4 &=& \sum_a N_a 2 m_1^a m_2^a \label{eq:fzerotadone}\\
24 &=& \sum_a N_a \left(2 (m_0^a)^2 + 2 m_0^a m_2^a\right)
 \label{eq:fzerotadtwo}\\
24 &=& \sum_a N_a \left(2 (m_0^a)^2 + 2 m_0^a m_1^a\right) 
\label{eq:fzerotadthree}
\end{eqnarray}
and the supersymmetry conditions become
\begin{eqnarray}
3 (8j_0^2+4j_0j_1+4j_0j_2+2j_1j_2) m_0 + 
3 (2j_0^2 +2j_0j_2) m_1 +
3 (2j_0^2 +2j_0j_1) m_2 =\\
8m_0^3 + 6 m_0^2 m_1 +6 m_0^2 m_2 + 6 m_0 m_1 m_2 \nonumber
\end{eqnarray}
and
\begin{eqnarray}
\label{eq:ellfibSUSY-2}
8j_0^3 +6 j_0^2 j_1 +6 j_0^2 j_2 + 6 j_0 j_1 j_2 >\hfill\\
3 (8m_0^2+4m_0m_1+4m_0m_2+2m_1m_2) j^0 + 
3 (2m_0^2 +2m_0m_2) j^1 +
3 (2m_0^2 +2m_0m_1) j^2 \nonumber.
\end{eqnarray}

We note for comparison with the torus discussion that, while naively
the total tadpoles are not all positive, if we add to
\eq{fzerotadone} some linear combination of the other equations,
the total tadpoles will become positive.  However one might
ask what content this has and whether there is any preferred choice
of basis for these equations.  We will answer this question shortly.

Many other elliptic fibrations can be found in the physics literature, as
they find application in F theory, in heterotic string compactification,
and so on.

\subsubsection{Toric hypersurface Calabi-Yau manifolds}

Such a CY is defined as the zero locus of a single equation in
a four-dimensional toric variety, the ``ambient'' space $T$.
Such a compact toric variety is determined by a four-dimensional polytope,
and the Calabi-Yau condition then reduces to the condition that
the polytope be reflexive.

A complete database of such polytopes is available at \cite{KreuzerSkarke}.
There are simple combinatoric algorithms for working out $c_2(M)$,
the cubic intersection form and the K\"ahler and Mori cones
from the polytope, see for example the appendix to \cite{DDF}.

It will be quite interesting to scan over this database and
work out the general pattern of solutions.  For example, does the
number of solutions grow with $b_2$, as one might expect?  Or
do the tadpole conditions become sufficiently complicated that
explicit solutions are difficult to find?

\subsection{Positivity and brane classes}

As in subsection \ref{sec:types}, certain basic consequences of the
supersymmetry and tadpole conditions can be analyzed {\it a priori},
leading to a division of the supersymmetric branes into three types
generalizing what we had for the torus.

We start by asking what it means for contributions to the tadpole
cancellation conditions \eq{gen-tadpole} to be positive or negative.
A natural definition to make is that 
the ``positive'' tadpoles are those in the Mori cone, {\it i.e.}
$[B]=r+T$ with $r\ge 0$ and $T\in MC(M)$.  This reduces to our previous
definition for the torus.  It is also known that
$$
ch_2(M) \in MC(M) 
$$
from Bogomolov's theorem applied to the tangent bundle $TM$.
Thus, we can again try to argue that solutions to the equations 
\eq{gen-tadpole} correspond to ways to decompose a positive vector into
a sum of positive and ``nearly positive'' contributions.  

For our example of a fibration over $F_0$, this definition tells us to
replace \eq{fzerotadone} with the constraint dual to the divisor
$D_0=B+2D_1+2D_2$.  This is a sum over the equations
Eqs. \ref{eq:fzerotadone}, \ref{eq:fzerotadtwo} and \ref{eq:fzerotadthree} 
with the corresponding coefficients, giving
\begin{equation}
92 = \sum_a N_a \left(
2 m_1^a m_2^a + 8 (m_0^a)^2 + 4 m_0^a m_1^a
+ 4 m_0^a m_2^a\right)
\end{equation}
While equivalent, the idea is that with this definition,
``most'' branes will make positive contributions to the right hand side.
For example, all D$5$'s wrapping holomorphic cycles clearly will do so,
by the definition of the Mori cone.

Now, unlike the torus and our example, a general Calabi-Yau manifold will
typically not have any preferred homology basis in which the K\"ahler
cone is simply defined by $J_i> 0\ \forall i$.  This is because the Mori
cone will typically have more than $b_2$ generators, and one must
enforce an inequality for each generator.

One can always take the basis for $H_2(M,\IZ)$ to consist of a subset
of the generators of the Mori cone, and we shall do this below.  In
this case, the K\"ahler cone will be the intersection of 
$J_i>0\ \forall i$ with additional inequalities.

We can now proceed to divide up the supersymmetric branes.

\begin{description}

\item {\Cb)} One nonvanishing tadpole.\\ 
Clearly the pure D$9$-brane
(with no flux) is a \Cb-type brane.  And, if we choose our basis for
$H_2(M,\IZ)$ to be a subset of the generators of the Mori cone, all
other vectors with a single tadpole will correspond to D$5$'s wrapping
holomorphic curves.  Thus, we have a complete set of ``filler''
branes, and can again argue that any configuration of \Ab\ and \Bb-branes
which undersaturates the tadpoles, can be completed in a unique way to
a configuration satisfying \eq{gen-tadpole}.  Furthermore, as we
discussed above, the \Cb-branes lead to no constraints on the K\"ahler
moduli, at least in the large volume limit.

Note that, in the typical case in which the Mori cone will has more
than $b_2$ generators, the precise identification of which branes are
the \Cb-branes will depend on our choice of basis for $H_2(M,\IZ)$.

\item {\Bb)} Branes whose tadpole contribution lies in the Mori cone:\\ 
This includes all supersymmetric D$5$-branes, which as before do not
constrain the K\"ahler moduli.  

It also includes a subset of the line bundles (magnetized D$9$-branes).
By definition, the tadpole contribution for a line bundle with flux $F$
will lie in the Mori cone if
\begin{equation}\label{eq:mori-F}
C(F,F,J) \le 0 \qquad \forall J\in KC(M)
\end{equation}
(the sign is correct, as can be checked for the torus; this generalizes
the minus signs we introduced in \eq{tadpoles}.

This is not an easy equation to satisfy, as the coefficients $c_{ijk}$
in our preferred basis will tend to be positive.  The \Bb-branes on the
torus are obtained by taking (say) $F_1>0$, $F_2<0$ and $F_3=0$, and
relying on the vanishing of all $c_{ijk}$ with coincident indices.
In other words, these are branes whose gauge fields are non-trivial in
only four of the six dimensions.

To illustrate the opposite case, consider a divisor $D_1$ for which 
the only non-zero intersections are $c_{111} > 0$ (for example a
local $\IP^2$.  Then \eq{mori-F} forces the corresponding $F$ to be zero.

\item {\Ab)} Branes whose tadpole contribution lies outside of 
the Mori cone:

This will include most line bundles.  One might go on to ask if there is
any sense in which their contributions to the tadpole equations
\eq{gen-tadpole} have a ``single negative component'' as was the case
for $T^6$.  This seems unlikely on the face of it, but fortunately need
not be the case for a finiteness argument to work.  

\end{description}

We should say the parallel between these definitions and those we made
for the torus is not complete; in particular the power law scalings 
discussed in subsection \ref{sec:summary} would not be expected to 
generalize.
What we do hope will generalize is the structure of the argument in
section 2 that the total number of configurations satisfying
\eq{gen-tadpole} is finite.  This is again clear at a single point in
Kahler moduli space, since the supersymmetry condition
\eq{im-gen-susy} forces an appropriate sum of the tadpole
contributions to be positive.  However we would like an argument which
does not use specific values of the K\"ahler moduli, and provides {\it
a priori} bounds.  This is an important goal for future work.

\subsection{Rough estimate of number of configurations}

Neglecting the supersymmetry conditions and just looking at the
general structure of the tadpole conditions \eq{gen-tadpole}, there is
an obvious guess for the growth of the number of brane configurations
as a function of the total scale of the tadpoles $T$.  We have to
count choices of the indices $m^a_i$ which satisfy
$b_2+1$ simultaneous quadratic conditions.  If we take these for a
single brane to all to
be of order $m$, a plausible first guess for the size of the tadpole
contributions is $m^2$, and thus we can take $m\sim\sqrt{T}$.  Since $m$
for a single brane has $b_2$ components, we can estimate
$$
\hat{\cal N} \sim T^{b_2/2}
$$
for a one-stack model, with possible higher power laws for 
multi-stack models.

This one-stack estimate would already lead to very large numbers for
typical Calabi-Yau manifolds with $b_2\sim 100$, comparable to numbers
of flux vacua.  Indeed, taken at face value, this estimate would be
far larger than the $(2\pi L)^{b_3}/(b_3)!$ estimate of \cite{ad}, as
it does not have the factorial in the denominator.  Fortunately, 
there are many possible sources of 
additional $b_2$ dependence, which could lead
to such factors.  In particular, in considering an explicit example in
\cite{DDF}, it was found that the multiplicity of terms in the
intersection form tends to make its value much larger (as $b_2^3$ in
the volume form), which could lead to a sizable suppression of the 
number, say to $(T/b_2^2)^{b_2/2}$.  It will be interesting to get
a better handle on this.

\section{Relevance for the problem of counting more general vacua}

Intersecting brane models are usually thought of as a special class of
constructions, useful for providing concrete examples, but not
necessarily mapping out the full set of possibilities; indeed one
might think they are a very small subset.

A contrary intuition comes from the realization that almost all
supersymmetric intersecting brane models contain many massless scalar
fields, coming from open strings stretched between pairs of branes.
Giving expectation values to these fields breaks part of the
four-dimensional gauge symmetry, and allows moving out into a larger
moduli space of configurations.  One might optimistically hope that
the general brane configuration could be obtained in this way.

Realizing this goal in practice requires a good deal of technical
control, in particular it requires knowing the world-volume
superpotential.  While this has been done for some noncompact
Calabi-Yau's (see for example \cite{Douglas:2000qw,Aspinwall:2005ur}),
for compact Calabi-Yau
manifolds it remains work in progress.

However, one might imagine that since the brane configurations are
special points in these moduli spaces with enhanced symmetry, knowing
all of them should give us some sort of global information about the
moduli space.  In this section we would like to propose an idea in
this direction.

What distinguishes the IBM brane configurations, is that they transform
simply under the symmetries of the compactification
manifold.  This is clear for hyperplanes in $T^6$, which admit an
action of the group of translations on $T^6$.  In the IIb picture, the
supersymmetric branes correspond to constant curvature connections on
line bundles over $T^6$, for which a translation acts as a gauge
transformation.  More generally, this condition defines the
equivariant bundles, the class of branes we suggested earlier as a
good IIb analog for general toric hypersurface CY's of the branes
usually used in constructing IBM's.

What use can we make of this observation?  Let us consider the problem
of counting all of the stabilized vacua in such a theory, which are
distinguished by the values of the open string moduli.  To do this, we
would need not just the tree level world-volume superpotential, but
the full superpotential with all non-perturbative corrections.  We
would then count its critical points $D_iW=0$.

As discussed in \cite{stat,ad,DSZ}, the number of critical points of the
superpotential is quasi-topological, in the following sense.  One can 
construct an index which weights critical points with a sign, which on
a compact moduli space would be the integral of a topological density,
\begin{equation}\label{eq:romega}
{\rm index} = \int_{\cal M} \det(R+\omega).
\end{equation}
The same integral makes sense on a non-compact moduli space as well,
and provides a good estimate for the number of vacua for a generic
superpotential.

How would one compute this?  Somewhat analogous problems arise in the
computation of instanton amplitudes in supersymmetric gauge theory
\cite{Dorey:2002ik,Nekrasov:2002qd} and in other topological
field theories.  In gauge theory, the $k$-instanton contribution to a
``protected'' coupling is an integral over $k$-instanton moduli space
of a corresponding form, typically the volume form or the Euler form
depending on the amount of supersymmetry.  

Now, while instanton moduli spaces are complicated, the essential
point which allows computing such integrals is that they and the
integrands admit continuous symmetries generated by Hamiltonians,
which allows the use of localization.  The obvious symmetry at hand in
gauge theory on Euclidean four-dimensional space is the lift of the
$SO(4)$ rotational symmetry -- since a rotation on a gauge field
configuration takes it into an {\it a priori} different gauge field
configuration, we get an action involving the corresponding points in
moduli space.  The Hamiltonian nature of the action turns out to
follow from that in Euclidean space in an analogous way.

The simplest localization arguments (which sufficed for
\cite{Nekrasov:2002qd}) use only an abelian group action, say 
$U(1)^2\subset SO(4)$.  Without going into details, the simplest
application of this type of argument 
is the Duistermaat-Heckman formula, according to which the integral 
over an $n$-dimensional space $\CM$ 
\begin{equation}\label{eq:DHform}
\int_{\CM} \omega^n e^{it H}
 = t^{-n/2} \sum_{{\rm fixed\ points}} \frac{1}{\det S''} ,
\end{equation}
where $H$ is the Hamiltonian generating the $U(1)$ symmetry, and
$\det S''$ is the one-loop determinant obtained by expanding the
integrand around a fixed point of the $U(1)$ action.  

The point is that, in favorable cases, the integral can be
reduced to a sum over contributions from the fixed points of the
$U(1)$ symmetry.  On the heuristic level of our discussion, one can
see various reasons why this might be applicable to the problem of
counting vacua.  For example, we might take $H$ itself as a model
potential $V=H$.  The fixed points of the isometry $\partial H=0$ would
then be exactly the ``vacua'' $V'=0$ we are trying to count.  We then
appeal to the quasi-topological nature of \eq{romega}, to argue that
the index counting vacua is the same for a more generic choice of 
potential.

Now, let us return to the problem at hand, of computing the number
of stabilized vacua which can be found by starting with IBM's and
varying open string moduli.  As we discussed, these will be solutions
of $DW=0$ for a generic superpotential on the moduli
space $\CM$ of holomorphic bundles (with specified Chern classes)
on the Calabi-Yau manifold $M$.  Since $M$ is K\"ahler, this moduli
space will be K\"ahler and thus symplectic.  And by general arguments,
the index counting critical points $DW=0$ will be 
the integral of the density \eq{romega} over this moduli space.

Now, suppose for a moment that $M$ admitted a continuous symmetry;
then so would the moduli space $\CM$.  In particular, for $M=T^6$,
translations will act on $\CM$, and the equivariant bundles will be
the fixed points.  As we discussed in section 5, these are just the
type IIB description of the IBM's we have been counting.

While a more general Calabi-Yau manifold will not have continuous
symmetries, if we restrict attention to the subset of bundles which
restrict from an ambient toric variety, the moduli space of such
bundles will again admit a continuous symmetry, and again the
equivariant bundles will be the fixed points.  These include sums of
line bundles and thus the right hand side of a formula like
\eq{DHform} will include a sum over the IBM's defined in section 5, as
well as other IBM's constructed using higher rank equivariant vector
bundles.  If these were fewer or comparable in number, and the
$1/\det$ factors of order unity, the sum could be estimated by
simply counting IBM's.

The main point of these heuristic arguments is that the IBM's are not
arbitrary points in moduli space, but special points which carry
information about the global topology of the moduli space.  While
there would be many technicalities to address in trying to make such
arguments precise, such as the presence of singularities, the 
computation of the $1/\det$ factors, and so on,
they suggest that the crude estimate 
``number of vacua roughly equals number of IBM configurations'' might be
reasonable for the open string sector.

\section{Conclusions}

In this paper we have developed tools for systematically analyzing
intersecting brane models.  We have looked in detail at a particular
toroidal orientifold and characterized the gauge groups and
intersection numbers associated with intersecting brane models on this
space.  

We have proven that there are a finite number of solutions of
the Diophantine equations arising from the supersymmetry constraint,
so that there are a finite number of different intersecting brane
constructions in this model.  We have developed estimates for the
number of configurations which contain a given gauge group.  The
number of models with a fixed gauge group scales polynomially in the
tadpole constraint, and is suppressed by a factor polynomial in the
sizes of the gauge group components.

We have looked at the distribution of intersection numbers which
govern the number of generations of chiral matter fields charged under
the gauge group living on the branes.  We found that generally the
intersection numbers are almost completely independently distributed,
and that the expected magnitude of the intersection numbers scales as
an inverse power of the size of the gauge group components.  We found
a mild suppression for prime intersection numbers and enhancement of
composite intersection numbers with many factors.

There are several ways to use the methods we have developed here.
Most concretely, one can use the estimates and algorithms we have
developed to first estimate the number of models with a given desired
gauge group and matter constant, and then to explicitly construct the
precise set of such models compatible with SUSY and other constraints
in an efficient way.  This gives a systematic way to build models
which agree with certain properties of the standard model, which can
then be studied for further physics predictions.  The standard choice
of the properties to reproduce are to get a gauge group containing the
standard model gauge group, three generations of matter, and two or
more Higgs doublets.  Additional charged matter is usually allowed,
but different works differ on the rules at this point.

The $T^6/\IZ_2\times\IZ_2$ orientifold we have focused on has been the
focus of much work of this type and thus we can test our results
against this work.  Most constructions leading to three generation
standard models require turning on discrete B fields, a class which we
have not enumerated.  However our estimates should apply to this case
and suggest that there should be about ten brane constructions which
give three generation Standard Models.  In fact, such models were
systematically constructed in \cite{Cveticll}, where 11 such models
were found.  Other less generic constructions using branes invariant
under the orientifold action led to 3-generation Pati-Salam models
with $U(4)\times U(2) \times U(2)$ gauge groups in
\cite{Cremades:2003qj,Marchesano:2004yq,Marchesano-Shiu-2}.\footnote{
See also \cite{Kokorelis:2002ip,Kokorelis:2002ns} which constructed
models with partial $N=1$ supersymmetry.}
We found
another pair of such explicit constructions here using our
understanding of the range of possibilities allowed.  It would be
worthwhile to understand the detailed phenomenology of these models
better, and to complete the search for standard model-like IBM's which
can be realized in this orientifold.

Using the techniques we have developed here, it would be
straightforward to automate such a search not only for the particular
orientifold we have considered here, but for a general toroidal
orientifold.  Carrying out a similar analysis for a broader class of
Calabi-Yau orientifolds involves further issues which we have
begun to address here, by setting out an approach to an analog of
the intersecting brane model construction for type I strings on
general Calabi-Yau manifolds.

Another way to think about the analysis of this paper is as a snapshot
of a piece of the full string landscape.  In \cite{Blumenhagen}
indications were found that some features of the standard model, such
as the components of the gauge group and the number of generations of
different types of matter, could be dialed fairly independently.  We
have developed a more complete picture of the distribution of
possibilities, and searched for correlations between distinct
intersection numbers using the information theoretic notion of mutual
entropy.  The results fit with the idea that different intersection
numbers are to a good approximation independently distributed, as suggested
in \cite{stat} and as emerged from the detailed studies of
\cite{Blumenhagen, Gmeiner}.

One definite correlation which does emerge is that larger rank gauge
groups come with much smaller intersection numbers, according to a
power law $I \sim ({\rm tadpole}/{\rm rank})^\alpha$ with a large
exponent $\alpha\sim 7$.

Of course, this is a very small sample even of type II orientifold
models, and we would suggest broadening the class of models (at least
to general Calabi-Yau orientifolds) before ascribing too much
significance to these results.  We should also keep in mind that the
simple statistical analyses we made can miss a great deal of
structure.  For example, anomaly cancellation fixes definite
relationships between the numbers of generations of different
types of matter which will be seen in the low energy theory.  Since
these conditions depend on the total matter content, they do not lead
to simple pairwise correlations.  Perhaps other, more subtle structure
of this type lies buried within the data.

As many have stressed, a noteworthy general feature of IBM's is that
their gauge groups tend to be larger than that of the Standard Model,
leading to new dynamics (if there is matter charged under both the SM
and the new gauge groups), and hidden sectors.  Similar observations
about heterotic string constructions, such as the presence of extra
$U(1)$'s and of course the second $E_8$ as a hidden sector, go back to
the mid-eighties (see \cite{Dienes:2006ut} for some recent statistical
work on heterotic constructions).  Clearly this would seem to be the
most promising area for making interesting phenomenological
predictions from this type of data.  As hypothetical examples, we
might imagine eventually making statements such as ``the vast majority
of known realistic string compactifications contain one or more extra
$U(1)$'s,'' or contain extra matter charged under the Standard Model
group, or hidden sectors with cosmological implications, or something
else.  Admittedly, as long as realistic models exist in which such a
property is not true, it would not be a ``Prediction'' whose
non-discovery would falsify string theory.  However, it would be a
``prediction'' that would be worth testing as we would know it fits
well with string theory; indeed to the extent that other vacuum
selection criteria (measure factors, anthropic selection or otherwise)
turn out not to depend on a property, it seems entirely reasonable to
say that a property realized in more vacua is favored by string
theory.  In addition, such a property might correlate with other observables;
conversely its non-discovery would be of great significance in
narrowing down the search for the correct vacuum.

There are many issues which must be addressed in order to make such
predictions.  On the more phenomenological side, of course one needs
additional input besides the charged matter content to decide what
part of the model is visible at accessible energies.  Even the
simplest case of matter which is chiral under the SM group (and thus
gets its mass from electroweak symmetry breaking), while generically
ruled out by existing data, is not absolutely excluded; these
constraints (which do not require any string theory input) have (to our
knowledge) not been spelled out completely in the literature.

On the more formal side, one has the problem that the number of
possible hidden sectors for a given visible sector model is
exponential in the difference between the tadpoles of the visible
sector and the total tadpole number.  This tends to make the problem
of enumerating even the possible matter sectors arising from all
string models intractable.  This motivates the style of analysis we
made, in which one enumerates models with the correct visible sector,
and settles for estimates for the number of possible hidden sectors.
However, to the extent that one must make a detailed model-by-model
study to determine which models are potentially consistent extensions
of the Standard Model, this is not very satisfying.  To resolve this
problem, we need simple and general criteria which tell us whether a
model is likely to work (or what degree of tuning is required to make
it work), without using detailed information.  

As an example which illustrates what we have in mind, it would be very
useful to have a simple criterion for a gauge sector to dynamically
break supersymmetry.  This has been much discussed in the literature,
beginning with \cite{Affleck:1984xz},
under a ``strong'' definition of supersymmetry breaking which
requires a theory not to have supersymmetric vacua.  Another
definition, which fits much better with the landscape picture, is that
any sufficiently long-lived metastable supersymmetry breaking vacuum
is acceptable.  Since this can be accomplished by tuning the
potential, one is not looking for a criterion which provides a
definite ``yes/no'' answer, but instead tells us what fraction of the
theories in a certain ensemble are expected to break supersymmetry,
and with what distribution of breaking scales.  This type of question
was addressed for IIb flux vacua in \cite{Denef:2004cf}; work such as
\cite{Dine:1995ag,Dimopoulos:1997ww} and especially the recent
\cite{Intriligator:2006dd} suggest that the time is ripe to address
this question in gauge theory.

Another important issue which we have not addressed in this paper is the
interplay between the open string sector and the closed string sector,
and the stabilization of moduli using fluxes.  In the vacua we have
considered, there are both open string and closed string moduli.
Because of the open string moduli, the intersecting brane
configurations we have constructed are just isolated points in a
continuous moduli space.  These points are rather special points with
enhanced symmetry, and as we have discussed they may give some
characteristic understanding of the full moduli space, but in a
complete theory we should stabilize these moduli.  This will naturally
occur when the closed string moduli are stabilized by the inclusion of
fluxes.  While some literature, such as
\cite{Blumenhagen:2003vr,Cascales-Uranga,%
Marchesano-Shiu-2,Cvetic:2004xx,Cvetic:2005bn},
has begun to 
address the problem of stabilizing open string moduli in parallel with
the stabilization of the closed string moduli by fluxes, a complete
mathematical characterization of this problem is not yet available.
The correct way to address this problem is to give a global
characterization of the open string moduli space in combination with
the closed string moduli space.  Just as fluxes induce a
superpotential on the closed string moduli space, which stabilizes the
closed string moduli as in \cite{Giddings:2001yu}, the same fluxes
will induce a superpotential on the open string moduli space.  A local
description of this superpotential on the open string moduli space was
given in \cite{Gomis:2005wc}.  An important open problem is to develop
a global description of this superpotential and tools for computing
supersymmetric solutions of the resulting equations of motion in which
both open string and closed string moduli are fixed.  This would give
a true characterization of a class of string vacua which could be
analyzed for standard model-like properties.

\vskip 0.2in {\it Acknowledgements}

We would like to thank Mirjam Cvetic, Bogdan Florea, Shamit Kachru,
Gordy Kane, Robert Karp, Maximilian Kreuzer, Jason Kumar, Paul
Langacker, John McGreevy, Nathan Seiberg, Steve Shenker, Gary Shiu,
Eva Silverstein, Scott Thomas, Jay Wacker and James Wells for helpful
conversations.  We would like to think the KITP for support and
hospitality during the initial stages of this work.  This research was
supported in part by the National Science Foundation under Grant
No. PHY99-07949.  The research of MRD was partially supported by DOE
grant DE-FG02-96ER40959.  The research of WT was supported in part by
the DOE under contract \#DE-FC02-94ER40818, and in part by the
Stanford Institute for Theoretical Physics (SITP).  WT would also like
to thank Harvard University for hospitality during part of this work.

\section*{Appendix A}

In this appendix we briefly review the appearance of the $\zeta$
function when summing over relatively prime integers.  The zeta
function $\zeta (s)$ is defined as
\begin{equation}
\zeta (s) = \sum_{n = 1}^{ \infty}  \frac{1}{n^s}  \,.
\end{equation}
This sum can be rewritten as a product over primes
\begin{equation}
\zeta (s) = \prod_{p \; {\rm prime}}
\frac{1}{(1-1/p^s)}  \,.
\end{equation}
The Euler totient function $\phi (n)$ gives the number of integers
less than $n$ which are relatively prime to $n$.  For a power of a
prime $p^k$,  it is easy to see that 
\begin{equation}
\phi (p^k) = \frac{p-1}{p}  p^k = (p-1) p^{k-1} \,.
\end{equation}
Similarly, if $n$ has distinct prime factors $p_i$ (which may be
repeated), then
\begin{equation}
\phi (n) = n\cdot\prod_{i}\frac{p_i-1}{p_i}  \,.
\end{equation}
As a result, we have
\begin{equation}
\sum_{n}\frac{\phi (n)}{n^s}  =
\prod_{p}\left[1 + \left( \frac{p-1}{p}  \right)
\left(p^{-(s-1)} + p^{-2(s-1)} + \cdots \right) \right]
= \prod_{p}\left[ \frac{1-p^{-s}}{ 1-p^{-s+1}}  \right]
= \frac{\zeta (s-1)}{ \zeta (s)} 
\label{eq:zeta-identity}
\end{equation}

\section*{Appendix B}

As discussed in section 4,
we list the 17 possible hidden sector \Ab-type branes which when
combined with a stack of 4 \Ab-type branes of the form
(\ref{eq:first-a}) and C branes with $R = 1$ and $S = 1$ give a
supersymmetric configuration which undersaturates all tadpoles.
\begin{eqnarray}
n = (6, 1, 3),\;\;\;\;\; m = (1, -1, -2) & \;\;\;\;\; &
(P, Q, R, S) =  (18, -12, 2, 3), \\
n = (5, 1, 3),\;\;\;\;\; m = (1, -1, -2) & \;\;\;\;\; &
(P, Q, R, S) =  (15, -10, 2, 3), \nonumber\\
n = (4, 1, 3),\;\;\;\;\; m = (1, -1, -2) & \;\;\;\;\; &
(P, Q, R, S) =  (12, -8, 2, 3), \nonumber\\
n = (3, 1, 3),\;\;\;\;\; m = (1, -1, -2) & \;\;\;\;\; &
(P, Q, R, S) =  (9, -6, 2, 3), \nonumber\\
n = (2, 1, 3),\;\;\;\;\; m = (1, -1, -2) & \;\;\;\;\; &
(P, Q, R, S) =  (6, -4, 2, 3), \nonumber\\
n = (6, 1, 3),\;\;\;\;\; m = (1, -1, -1) & \;\;\;\;\; &
(P, Q, R, S) =  (18, -6, 1, 3), \nonumber\\
n = (5, 1, 3),\;\;\;\;\; m = (1, -1, -1) & \;\;\;\;\; &
(P, Q, R, S) =  (15, -5, 1, 3), \nonumber\\
n = (4, 1, 3),\;\;\;\;\; m = (1, -1, -1) & \;\;\;\;\; &
(P, Q, R, S) =  (12, -4, 1, 3), \nonumber\\
n = (5, 2, 2),\;\;\;\;\; m = (1, -1, -1) & \;\;\;\;\; &
(P, Q, R, S) =  (20, -5, 2, 2), \nonumber\\
n = (4, 2, 2),\;\;\;\;\; m = (1, -1, -1) & \;\;\;\;\; &
(P, Q, R, S) =  (16, -4, 2, 2), \nonumber\\
n = (10, 1, 2),\;\;\;\;\; m = (1, -1, -1) & \;\;\;\;\; &
(P, Q, R, S) =  (20, -10, 1, 2), \nonumber\\
n = (9, 1, 2),\;\;\;\;\; m = (1, -1, -1) & \;\;\;\;\; &
(P, Q, R, S) =  (18, -9, 1, 2), \nonumber\\
n = (8, 1, 2),\;\;\;\;\; m = (1, -1, -1) & \;\;\;\;\; &
(P, Q, R, S) =  (16, -8, 1, 2), \nonumber\\
n = (7, 1, 2),\;\;\;\;\; m = (1, -1, -1) & \;\;\;\;\; &
(P, Q, R, S) =  (14, -7, 1, 2), \nonumber\\
n = (6, 1, 2),\;\;\;\;\; m = (1, -1, -1) & \;\;\;\;\; &
(P, Q, R, S) =  (12, -6, 1, 2), \nonumber\\
n = (5, 1, 2),\;\;\;\;\; m = (1, -1, -1) & \;\;\;\;\; &
(P, Q, R, S) =  (10, -5, 1, 2), \nonumber\\
n = (4, 1, 2),\;\;\;\;\; m = (1, -1, -1) & \;\;\;\;\; &
(P, Q, R, S) =  (8, -4, 1, 2) \nonumber
\end{eqnarray}

\end{document}